\documentclass[12pt,preprint]{aastex}

\def\msun{M$_{\odot}$}

\def\etal{et al.}			
\def\H0{$H_0$~= 75 \kms\ Mpc$^{-1}$}

\def\kms{km s$^{-1}$}

\def\ref{\par\noindent\hangindent 30pt}

\def\v16{$\Delta V_{1-6}$}
\def\lea{\mathrel{<\kern-1.0em\lower0.9ex\hbox{$\sim$}}}
\def\gea{\mathrel{>\kern-1.0em\lower0.9ex\hbox{$\sim$}}}

\received{2006 February 13}

\begin{document}

\title{STAR CLUSTER DEMOGRAPHICS. I. A GENERAL FRAMEWORK AND
APPLICATION TO
THE ANTENNAE GALAXIES\altaffilmark{1} }

\author{\sc Bradley C. Whitmore\altaffilmark{2},
Rupali~Chandar\altaffilmark{3,4}, and  S. Michael Fall\altaffilmark{2}}
\affil{email: whitmore@stsci.edu, rchandar@ociw.edu, fall@stsci.edu}

\altaffiltext{1}{Based on observations with the NASA/ESA {\it Hubble
Space Telescope}, obtained at the Space Telescope Science Institute,
which
is operated by the Association of Universities for Research in
Astronomy,
Inc. under NASA contract NAS5-26555. }
\altaffiltext{2}{Space Telescope Science Institute, 3700 San Martin
Drive,
Baltimore, Maryland 21218}
\altaffiltext{3}{Center for Astrophysical Sciences, Johns Hopkins University,
3400 N. Charles Street, Baltimore, MD 21218}
\altaffiltext{4}{Carnegie Observatories, 813 Santa Barbara St., Pasadena, CA 91101-1292}

\begin{abstract}

We present a framework for understanding the demographics of star
cluster systems, and develop a toy model which incorporates a
universal initial power law mass function, selected formation
histories, selected disruption laws, and a convolution with common
artifacts and selection effects found in observational data. A wide
variety of observations can be explained by this simple model,
including the observed correlation between the brightest young cluster
in a galaxy and the total number of young clusters. The model confirms
that this can be understood as a statistical size-of-sample effect
rather than a difference in the physical process responsible for
the formation of the clusters, suggesting that active mergers have the
brightest clusters simply because they have the most clusters.  A
comparison is made between different cluster disruption laws and it is
shown that the break in the $dN/d\tau$ diagram used to determine the
parameters in the Boutloukos \& Lamers model may be produced by
incompleteness near the breakpoint.
A model of the Antennae galaxies is developed and compared
with the observational data; this extends the mass-age range for cluster
comparison over previous studies.  An important component of our model
is the use of a ``two-stage'' disruption process, with a very high
``infant mortality'' rate for the clusters with ages less than
$\approx$ 10$^8$ yrs (i.e., roughly 80\%--90\% are lost each
factor of ten in time, $\tau$, independent of mass),
and two-body relaxation, which becomes the dominant disruption
mechanism at older ages, preferentially removing the lower mass
clusters.  Hence, in our model, stars from the dissolved clusters form
the field population. We note that a 90 \% infant mortality rate for
each factor of ten in  $\tau$ (i.e., $dN/d\tau$ $\propto$
$\tau^{-1}$) is consistent with all measured young cluster populations,
including those in the Antennae, Small Magellanic Cloud, and the
Milky Way.
In fact, the population of clusters in the Antennae can be viewed as a
scaled-up version of the Milky Way in many respects, with a scale
factor of roughly 1000 times the Lada \& Lada sample of embedded star
clusters in the local Milky Way.  We find no evidence for a truncation
of the Antennae cluster mass function at the high mass end.

\end{abstract}

\keywords{galaxies: star clusters, galaxies: interactions, galaxies:
individual (NGC 4038/4039)}


\section{Introduction}

The discovery of young massive clusters (hereafter YMCs) in merging
and starburst galaxies has been a major catalyst for the field of
extragalactic star clusters in recent years (see Whitmore 2003 and
Larsen 2005 for reviews). In particular, the fact that the most
massive of these young clusters have all the attributes expected of
young globular clusters means that we can study the formation of
globular clusters in the local universe rather than trying to
understand how they formed some 13 Gyr in the past.

A natural inference one might draw from the discovery of YMCs in
merging and starburst galaxies is that some ``special physics'' is
occuring in these environments; that normal quiescent spirals can only
form low mass open clusters.  However, the discovery of a small number
of YMCs in normal spiral galaxies by Larsen and Richtler (1999) showed
that this inference is incorrect. The formation mechanism required to
make YMCs appears to be more ``universal'' than originally
thought. Early hints that this phenomenon might be quite widespread
were the discovery of YMCs in barred galaxies (Barth, Ho \& Filippenko
1995), M33 (e.g., Chandar, Bianchi, \& Ford 1999a,b), and the central region
of our own Milky Way Galaxy (Figer \etal\/ 1999).  The fact that
most studies of the luminosity functions of young compact
clusters have found them to be power laws of the form $\phi(L)dL
\propto L^{\alpha} dL$ with a value of $\alpha$ $\approx$ --2 (e.g.,
compilations in Whitmore 2003, Larsen 2005, 2006; although see \S 4.2 
for alternative views)
provides additional evidence for the
universality hypothesis.  The main difference appears to be the
normalization of the power law, since young active mergers typically
have thousands of YMCs, older mergers and starbursts have hundreds of
YMCs, and normal spiral galaxies have between zero and one hundred YMCs.

While cluster mass functions have been determined for only a small
subset of these galaxies (due to the need for cluster age estimates
which are observationally expensive), a similar power law form
has been found for the mass and luminosity functions in the Antennae,
(i.e., $\psi(M)dM \propto M^{\beta} dM$ with a value of $\beta$
$\approx$ --2; Zhang \& Fall 1999; Fall 2004; Fall, Chandar, \&
Whitmore 2006), in the Small and Large Magellanic Clouds
(Hunter \etal\/ 2003), as well as a few other galaxies
(see review by Larsen 2006). 

Since the number of galaxies with sufficient data to form even a
meaningful luminosity function was small, Whitmore (2003; originally
presented in 2000 as astro-ph/0012546) enlarged his sample by
comparing the luminosity of the brightest young cluster in a galaxy
with the number of clusters in the system brighter than M$_V$ =
--9. Figure 1 shows an updated version of this figure, with the
primary improvements being the addition of galaxies from Tables 1$-$5
of Larsen (2005) and updated cluster numbers for several
galaxies. Although the galaxy sample is heterogeneous, this diagram
shows no evidence for a discontinuity in the ability of quiescent
spiral galaxies to form bright star clusters, when compared with
mergers and starbursts. The observed slope is consistent with the
hypothesis that all of the galaxies have the same universal cluster
luminosity function; a power law with index $\approx-2$. This suggests
that active mergers have the brightest clusters only because they have
the most clusters (i.e., a size-of-sample effect).  {\it It appears to
be a matter of simple statistics rather than a difference in physical
formation mechanisms.}  A more detailed discussion of Figure 1 is
included in \S 3.1.

While the luminosity and mass functions of young star
clusters appear to be power laws, the luminosity and mass functions of
old globular clusters
are peaked, and are generally characterized with a mean value of M$_V$
$\approx$ $-7.4$ (corresponding to a mean mass $\approx$ 2 $\times$
10$^5$ \msun\/) and width $\sigma \approx 1.4$ mag (e.g., see
compilation in Ashman \& Zepf 1998). Several theoretical works (Fall
\& Rees 1977; Vesperini 1998; Fall \& Zhang 2001) have studied the
effects of two-body relaxation, tidal shocks, and stellar mass loss on
cluster populations. These studies have suggested that a natural way
to explain this apparent evolution is via the preferential
destruction of the fainter, less massive clusters, due primarily to
the effects of two-body relaxation in a tidal field.  Modelling of
this effect has shown preferential dissolution of lower mass clusters
hundreds of Myr after they form.  Over $\sim10$ Gyr this can introduce a
turnover in an initial power law mass distribution (e.g., Fall \&
Zhang 2001), similar to what is observed for ancient globular cluster
systems in a number of massive galaxies.

More recently, several studies (e.g., Fall 2004; Whitmore 2004; Fall,
Chandar \& Whitmore 2005; Bastian \etal\/ 2005; Mengel \etal\/ 2005)
have suggested that the majority of stellar clusters are disrupted
within the first $\sim10$~Myr of life, and hence are presumably not
bound to begin with.  More specifically, Fall \etal\/ (2005) find that
$dN/d\tau \propto \tau^{-1}$ for clusters with ages $\lea$100~Myr and
masses $\gea3\times10^4~M_{\odot}$ in the Antennae galaxies, implying
that roughly 90 \% of the clusters are disrupted each decade (i.e.,
factor of ten) of time, $\tau$.  This rapid destruction, or ``infant
mortality'', is the dominant contributor to cluster demographics over
this period of time. A similar effect has been noted for
$10^2-10^3~M_{\odot}$ embedded clusters in the Milky Way (Lada \& Lada
2003).  In apparent contradiction to these results for the young
cluster systems in the Antennae and Milky Way, Rafelski \& Zaritsky
(2005) suggest $dN/d\tau \propto \tau^{-2}$ for the cluster population
in the Small Magellanic Cloud.  However, a re-analysis of their data
(Chandar \etal\/ 2006) indicates that $dN/d\tau \propto \tau^{-1}$ in
the SMC as well (see \S4.3 and Figure~15).

As outlined above, previous results on the Antennae cluster system
have provided some of the motivation for the framework presented in
the current paper.  These have revealed that the mass distribution is
approximately independent of age (Zhang \& Fall 1999, Fig. 3) and that
the age distribution is approximately independent of mass for massive
young clusters (Fall et al. 2005, Fig. 2).  As in these previous
works, we define clusters as the compact, barely-resolved objects
detected in galaxies such as the Antennae (e.g., Whitmore \etal\/
1999).  Since we cannot directly assess whether a given aggregate of
stars is bound or unbound based on the photometric data alone, we make
no assumption about how long a cluster might survive.  The development
of the model presented here allows us to extrapolate outside of the
ranges used previously to establish the cluster mass, age, and
luminosity relations, and simultaneously to compare model predictions
with the observed properties of the host galaxy.  To the degree that
our models fit the integrated properties of a galaxy (e.g., the total
stellar mass and total luminosity) we can have some confidence that
these extrapolations are reasonable.

In summary, it appears plausible that most star clusters form from a
universal initial cluster mass function, which is then modified by 
several disruption  mechanisms.   The current paper describes a  
simple toy
model designed to test this framework in detail. In order to
facilitate direct comparison with observations, common selection
effects and data reduction artifacts are included in the model. The
results are then compared with observations of the Antennae cluster
system which extend the mass-age domain explored previously.
An earlier version of some of this work was described in Whitmore
(2004). Paper 2 (Chandar \& Whitmore 2006, {\it in preparation}) in
this series will extend the comparison to a number of other nearby
galaxies, and introduce an objective, 3-dimensional classification
system that helps facilitate the comparison (see Whitmore 2004).  
We are therefore testing a specific view, that most star formation occurs 
in
groups and clusters of stars according to a single power law mass
function. This distribution is subsequently modified, with disrupted
clusters forming the field. An alternative view is that clusters form
with a Gaussian mass function (e.g., Mengel \etal\/ 2005;  
Gieles \etal\/
2006; Fritze-van Alvensleben 2004). We discuss the evidence for
and against this possibility in \S 4.2.

This paper is organized as follows: \S 2 develops the framework for
understanding the observed properties of star cluster populations; \S
3 presents some results and predictions of the model, in particular a
simulation of the M$_V$(brightest) vs. log(N) diagram, and a
comparison with observations of the Antennae galaxies; and \S 4
discusses broader implications and some potential anomalies for the
universal framework.  Finally, in \S 5 we summarize the main
conclusions of this work.

\section{A Universal Framework for Understanding Star Cluster
Demographics}

In this section we develop a framework for understanding the
demographics of star clusters, following the theme outlined in the
Introduction. A simple toy model has been created using the IDL
programming language to implement this framework.  The key ingredients
of the model are:

\noindent{\bf{A. Initial Mass Function}}
\begin{itemize}
\item assumed to be a powerlaw ($\psi(M)dM \propto M^{\beta}dM$) with
index $\beta=-2$. The effects of using
other indexes are briefly examined in \S 3.1 and \S 3.2.
\end{itemize}

\medskip

\noindent{\bf{B. Various Cluster Formation Histories\\}} Two cluster
formation histories are included; a constant rate of
formation and a Gaussian burst.  Different components can be
combined together as needed to produce the model for a galaxy. For
example, the 5-component model for the Antennae cluster system
discussed in \S 3.2 consists of three constant-formation components
and two Gaussian components.  All star formation is assumed to occur
in clusters. The field stars result from the disruption of the
clusters. The Bruzual \& Charlot (2003) models are used for the
spectral energy distributions [SEDs].
\begin{itemize}
\item constant formation in time ($dN/d\tau=$constant)
\item Gaussian burst ($dN/d\tau \propto
\tau_{0}\mbox{e}^{-\Delta\tau^{2}/2\sigma^2}$)
\end{itemize}
\medskip

\noindent{\bf{C. Various Disruption Laws\\}} Four
disruption laws are included.
It is assumed that the stars from the disrupted clusters become the
field stars in the galaxy.

\begin{itemize}

\item constant mass loss ($M=M_{0} - \mu_{ev}\tau$).
This is the time dependence resulting from two-body relaxation for a
cluster in the tidal field of the host galaxy (i.e., a value of
$\mu_{ev} \sim 2\times10^{-5}~M_{\odot}~\mbox{yr}^{-1}$ matches the
detailed calculations in Fall \& Zhang 2001; see their Figures~1 \&
4, and their equation 6). We also note that observations of faint globular 
clusters in the Milky Way (see Fall \& Zhang 2001, Figure 3), and M87
(Waters \etal\/ 2006) agree with this  functional form.

\item constant number loss ($dN/d\tau \propto \tau^{\gamma}$).  A
value of $\gamma$ = $-1$ results in 90 \% loss each decade of $\tau$;
this is defined as 90\% ``infant mortality''.

\item a two-stage disruption model,
   incorporating constant number loss (infant mortality)
for the first 10$^8$ years and constant mass loss (two-body
relaxation) at all ages.

\item an empirical disruption law from Boutloukos \& Lamers (2003)

\noindent ($t_{dis}(M_{cl}) = t_{4}^{dis}
(M_{cl}/10^4~M_{\odot})^{\gamma_{BL}}$), where $t_4^{dis}$ is the
disruption time for a $10^4$~\msun\/ cluster and $\gamma_{BL}$ is the
mass dependence of the disruption timescale.

\end{itemize}
\medskip

\noindent{\bf{D. Convolution with Common Observational Artifacts and
Selection
Effects}}

Three artifacts and selection effects are currently
incorporated in the  model. \S2.5 describes the following
in more detail,
and outlines plans for the inclusion of additional selection effects
in the future.

\begin{itemize}
\item magnitude threshold
\item reddening and extinction
\item artifacts from age-dating algorithms
\end{itemize}
\medskip

As briefly mentioned in the Introduction, the mass and age
distributions for massive young clusters in the Antennae, which
provided some of the motivation for this work, are {\it independent}
of one another (Fall 2006).  This means that the joint distribution
$g(M,\tau)$ must be the product of the mass and age distributions,
$\psi(M)$ and $\chi(\tau)$, which roughly follow the relation
$g(M,\tau) \propto M^{\beta}\tau^{\gamma}$.  We have adopted this
approach, hence simplying our model.

We should note that these relationships are only meant to be
first-order approximations to reality, designed to develop a simple
framework that captures the dominant mechanisms required to understand
the demographics of star clusters in galaxies.
For example, Fall, Chandar \& Whitmore (2006) show that a value of
$\beta\sim-2$ is a reasonable overall fit to the data for the young
($\lea100$~Myr) Antennae cluster system over the mass range
$10^4$~M$_{\odot}$ to $10^6$~M$_{\odot}$.  The framework presented
here allows us to test whether a value of $\beta\sim-2$ beyond this
range is
consistent with other available data and our assumed model.

\subsection{Initial Mass Function}

The choice of a power law with index $\beta =$ --2 for the initial
distribution of mass has already been motivated in the Introduction.
We address recent claims that the initial cluster mass function in
some nearby galaxies deviates
significantly from a power law in \S 4.2.  One nice attribute of
adopting $\beta$ = --2 is that the number of clusters in each decade
of decreasing mass increases by a factor of 10, hence there is equal
mass in each decade (e.g., the total mass of all the clusters in the
range $10^5 - 10^6$ \msun\/ is the same as the total mass of all the
clusters in the range $10^3 - 10^4$ \msun\/).

A word about boundary conditions is in order, since integrating a
power law mass function with an index of $-2$ over all mass ranges would
lead to an infinite mass.  The lower mass limit is controlled by the
masses of individual stars, roughly 0.1 $-$ 100 \msun\/.  The highest
mass
cluster formed may be controlled by statistics, namely the very low
probability of producing a cluster with more than 10$^7$ \msun\/, or
possibly by physics if there is a physical upper limit to the allowed
mass for an individual cluster.
Therefore, there are  $\approx$ $5-8$
decades (i.e., factors of ten) 
of cluster masses to be concerned with.  Hence, if the
majority of
stars are born in star clusters, each decade in initial mass of the
clusters should produce roughly 10$-$20 \% of the total mass in stars
for a
given galaxy.

One may argue that the mass of stars in globular clusters in the Milky
Way is only about 10$^{-4}$ of the total mass of the Galaxy halo,
rather than the $\approx$ 40$-$60 \% expected from the argument above
and
the fact that globular clusters span roughly three decades in mass
(from $\approx$ 10$^4$ to $\approx$ 10$^7$ \msun\/).
However, as we shall
see, this is consistent with the dominating role that cluster
disruption plays in determining the demographics of clusters. Hence,
most of the stars initially formed in clusters eventually end up in
the field.

We also note that our observations in galaxies like the Antennae only
allow us to determine directly the initial mass function over about
two decades of mass, from $\approx$ 10$^{4}$ to 10$^{6}$
\msun\/. However, to the degree that our models fit the integrated
properties of a galaxy (e.g., the total stellar mass and total
luminosity) we can have some confidence that extrapolations outside of
these ranges are reasonable.

\subsection{Cluster Formation Histories }

Our model accomodates linear combinations of
constant cluster formation and Gaussian bursts to reproduce
observations of different galaxies.
For example, typical spiral galaxies
can be modeled by an initial burst of star formation to produce the
ancient GC population roughly 13~Gyr ago, combined with a constant
rate of
formation since that time.

We begin our examination of the cluster formation history with perhaps
the simplest model; constant formation over the past 13 Gyr with
no cluster disruption.  Figure 2a (upper panel) shows the
log($M/M_{\odot}$) vs. log$\tau$ diagram for this model, using our
standard initial mass function which has a power law with index
$\beta$ = --2.0.

We note that, similar to our choice of $-2$ for the power law index
for the mass distribution, the choice of a constant formation rate
simplifies the accounting, since there are a factor of 10 more
clusters in each decade of time, by definition.  Similarly, the
boundary conditions in the temporal dimension are fixed, since
clusters (and the brightest stars within them) require $\approx$ 1 Myr
to form. Hence, there are only about 4 decades in $\tau$ to integrate
over (i.e., $\sim$1 Myr to $\sim$10 Gyr).

Observations of cluster systems are limited by the quality of the
data.  While cluster detection is a complicated function of object
brightness, local background level and complexity, and object size,
the most basic observable parameter which sets the limit for a given
dataset is the brightness of an object (i.e., the magnitude threshold).
Figure 2b (middle panel)
shows the same simulation as Figure 2a, but with a magnitude threshold
imposed by restricting the simulated sample to clusters with
M$_{V}$  $\leq$ $-9$.  A comparison of Figures 2a
and 2b demonstrates the importance of the magnitude threshold. For
example, if this selection effect is not considered, one might
conclude that
the disruption of faint clusters is the reason for the lack of
objects in
the bottom right of Figure 2b.  This is a cautionary reminder that we
must
be careful not to confuse evidence for disruption with artifacts
caused by observational selection effects.  We will return to this
point in
\S2.3.

Figures 2a and 2b also provide a good opportunity for demonstrating
the size-of-sample effect. Taken at face value, this figure suggests
that massive clusters were only produced in the distant past, since
the most massive cluster with an age less than 10 Myr is $\approx$
10$^4$ \msun.  However, the clusters {\it were all taken from the same
power law mass function}; the only reason for the lack of
massive young clusters is the factor of 1000 fewer clusters in the
$1-10$ Myr bin when compared with the $1-10$ Gyr bin.
Hence statistics, rather than physics, are responsible for the lack
of massive young clusters in Figure~2. There are simply not enough
young clusters to produce the 2 $\sigma$ statistical deviation
required to form a 10$^5$ \msun\/ cluster, let alone the 3 $\sigma$
deviation required for a 10$^6$ \msun\/ cluster.

Figure 2c (bottom panel) shows the same model as Figure 2b, but
plotting the luminosity (rather than the mass) versus age.  This
diagram is more directly related to observations, and serves as a
reminder that the magnitude threshold (defined by the lower edge seen
horizontally in Figure~2c) translates to a diagonal line in Figure 2b
due to the dimming of the clusters with age.

\subsection{Inclusion of Various Cluster Disruption Laws}

At present, four idealized disruption laws have been incorporated
into the toy model, as outlined in \S 2.  In Figure~3, we plot age
versus mass distributions for simulated clusters for three of the laws
(the fourth, the two-stage disruption model, is a combination of two
of the other laws, and will be discussed in more detail in \S 3.2).
Each model shown in Figure~3 begins with a constant rate of cluster
formation, which is subsequently modified by the various disruption
laws described below. Figure~4 shows the corresponding age
distributions for the twelve models presented in Figure~3.

In the constant {\it mass} loss models (shown in the left columns of
Figures~3 and 4), a fixed amount of mass is removed from each cluster
during each unit of time. A value of $\mu_{ev}\sim10^{-5}$ \msun\//yr
results in reasonable values (e.g., a 10$^4$ \msun\/ cluster is
disrupted in 10$^9$ years). The resulting mass distribution of
surviving clusters after a Hubble time looks similar to a typical
distribution of globular clusters, with a turnover around 10$^5$
\msun, as shown in Figure~5.  The constant mass loss disruption law
was motivated by the detailed calculations made in Fall \& Zhang
(2001), which shows that two-body relaxation is the
dominant destruction mechanism for clusters older than about 10$^8$
yrs, 
and that the mass loss of a cluster via two-body relaxation is
linear with age.  The peak in the mass distribution will also increase
linearly with time, as shown in the 10$^{-5}$ \msun\//yr panel of
Figure~3 (which mirrors the result in Fall \& Zhang 2001
Figures~3$-$11).  We note that two-body relaxation alone destroys very
few clusters in the first 10$^8$ yrs
(cf. Baumgardt \& Makino 2003), which is contrary to the
important role ``infant mortality'' plays in the evolution of cluster
systems. 

Models of constant {\it number} loss, or ``infant mortality'', reduce
the existing population of clusters by a fixed percentage (for e.g.,
0\%, 50\%, 80\%, or 90\% {\it each decade} in $\tau$; corresponding
to slopes of $\gamma$ = $0.0$, $-0.3$, $-0.7$, and $-1.0$ for the age
distribution).  This type of (mass-independent) cluster disruption
model is motivated by the age distribution of clusters observed
originally in the Antennae (as described in detail in Fall et
al. 2005, and further discussed in \S2.4).
The center column of panels in Figures~3 and 4 show illustrative
examples for different values of constant number loss. Over the first
$\sim10$~Myr in the life of a cluster, photoionization, stellar winds,
and supernovae can inject sufficient energy into the inter-cluster
medium to remove much of the gas.  This gas removal can unbind the
cluster (e.g., Hills 1980; Lada, Margulis \& Dearborn 1984; Boily \&
Kroupa 2003a,b; Fall \etal\/ 2005).  Following this stage, stellar
evolution will continue to significantly remove mass from the cluster
for longer periods of time (e.g., Applegate 1986; Chernoff \& Weinberg
1990; Fukushige \& Heggie 1995).

The middle column of Figure~3 shows that as the fraction of the disrupted
population is increased, the predicted ratio of young to old clusters
increases dramatically.  The age distribution for cluster populations
affected by different levels of infant mortality are shown in
Figure~4.  The top-middle panel shows that for a constant rate of
cluster formation, with {\it no} disruption, a mass limited cluster
sample would have a flat distribution in the $\mbox{log}~dN/d\tau$
vs. $\mbox{log}\tau$ diagram.  Higher values of infant mortality would
cause steeper age distributions, as shown by the solid
points in the next three (middle) panels in Figure~4.  The predicted
age distributions in Figure~4 are also compared with the empirical age
distribution for the Antennae galaxies (presented in Fall et
al. 2005).  In this figure, it is obvious that the 90\% constant number
loss (i.e., infant mortality-type disruption) gives the best match
between models and observations, although the 80\% model is also
reasonable given the uncertainties (e.g., the datapoint for the
youngest clusters may be artificially high due to a recent burst of
star formation in the ``overlap'' region; see \S 2.4).

The main shortcoming of this model is that we do not expect it to be
relevant past a few $\times $10$^8$ years, at which point two-body relaxation
should
begin to dominate (Fall \etal\/ 2005). We also note that from an
empirical standpoint, the 90 \% disruption rate has only been shown
to be roughly linear out to about this age (i.e., Figure 2 from Fall
\etal\/ 2005). After this point stellar contamination begins
to affect the dataset.

The third cluster disruption law (not shown) is a two-stage
disruption process, which includes both the constant number loss
(infant mortality) and the constant mass loss (two-body relaxation)
models.  While this two-stage process has been discussed previously
(e.g., Fall et al. 2005), this work provides the first detailed
application to cluster systems.
We note that the resulting disruption law is therefore {\it
mass-independent} for young ages and {\it mass-dependent} for older
ages.  This will be our primary tool when building our model for the
Antennae, and hence will be described in more detail in \S 3.2.

The fourth cluster disruption law is described in detail in
Boutloukos \& Lamers (2003; hereafter BL03). They fit the formula
$t_{dis}(M_{cl}) = t_{4}^{dis} (M_{cl}/10^4~M_{\odot})^{\gamma_{BL}}
$,
where $t_4^{dis}$ is the disruption time for a $10^4$~\msun\/ cluster
and $\gamma_{BL}$ is $\approx$ 0.6. The value of $t_4^{dis}$
is derived based on apparent bends in the log $dN/d\tau$
vs. log$\tau$ and 
diagram, which as we shall see, vary dramatically from galaxy to
galaxy.

Lamers, Gieles, \& Portegies Zwart (2005; hereafter LGPZ05) use this
technique to determine a very long characteristic disruption time
($t^{dis}_{4}\sim8\times10^9~\mbox{yr}$) for clusters in the SMC.  Our
simulations for LGPZ05 SMC-type disruption is shown in the upper right
panels of Figures~3 and 4.  Note that using the LGPZ05 value of the
disruption formula results in essentially {\it no} clusters with $\gea
10^4$ \msun\/ being disrupted in the SMC.  Conversely, LGPZ05 find a
very short disruption time for clusters in M51 ($t_{4}^{dis}=70$~Myr),
with essentially {\it all} clusters older than 1~Gyr being disrupted.
LGPZ05 interpret this large variation in $t^{dis}_{4}$ to be caused by
different disruption rates due to differences in the local ambient
density (for the clusters which have survived gas removal in the first
10~Myr).  However, the $t^{dis}_{4}$ derived for M51 clusters is
shorter by nearly an order of magnitude than predictions from N-body
simulations based on the estimated density of M51 (Lamers et
al. 2005).  Is it really possible that the disruption laws in these
two galaxies could be so different ?

The disruption time derived by LGPZ05 for M33 ($t^{dis}_{4} =
6\times10^8~\mbox{yr}$) is intermediate between the values given for
the SMC and M51. We note that the estimated values of $t^{dis}_{4}$
decrease monotonically as a function of distance for the SMC, M33, and
M51.  This led us to investigate whether the bend in the age and mass
diagrams that LGPZ05 use to derive $t^{dis}_{4}$ could be an artifact
caused by the magnitude threshold (which is typically distance
dependent), combined with various selection effects.

In Figure~6 we show the resulting age distributions for an artificial 
population of
clusters formed continuously and subject to a magnitude threshold,
with three different mass limits assumed.
If a sufficiently high mass limit is chosen, the resulting age
distribution is flat as expected, since it is not affected by
completeness or disruption of clusters.  The diagram also shows that
as lower mass limits are used, the resulting age distributions become
incomplete
due to the magnitude threshold, resulting in an apparent break.  {\it The
vertical lines show that the break in the age distribution occurs near
the age where the magnitude threshold intersects with the mass limit.}
Similar results are found using the infant mortality or two-stage
disruption
laws, but with sloping $dN/d\tau$ diagrams.

One prediction of this analysis
is that two galaxies at the same distance, observed using similar
magnitude thresholds, should result in similar estimates of
$t^{4}_{dis}$ (i.e., the observational artifacts would result in the
same position of the bend). This appears to be the case, since de
Grijs \& Anders (2006) find the same value of log~($t^{4}_{dis}$) for
the LMC ($9.9\pm0.1$) as LGPZ05 found for the SMC ($9.9\pm0.2$).  
We conclude that currently, there is insufficient data to construct a sample of
nearby galaxies which cover a range of distances and local densities
needed to definitively establish whether the derived $t^{4}_{dis}$ timescales
result from biases or density.  

To test our suspicion that this may be
a magnitude threshold bias, we used our Antennae dataset according to
the prescription of BL03.  When restricted to dynamic ranges of age
and mass where the data are approximately complete, no breaks are seen
in the age and mass distributions (i.e. they are smooth power laws).
If age and mass distributions are plotted over larger dynamic ranges
where we know completeness is an issue,
breaks are seen which imply even shorter $t^{4}_{dis}$ timescales than
found for M51.  Hence, in both this case, and in the artificial
cluster simulation described above, we know that the apparent
breaks are caused by selection effects and incompleteness. We predict
that as age and mass distributions become available for other galaxies
it will become evident that any apparent breaks are roughly dependent
on the depth of the cluster survey (i.e., the distance), and hence
caused by an artifact.

\subsection{The Need for Rapid Cluster Disruption (i.e., ``Infant
Mortality'')}

The age-luminosity distribution for the 
constant cluster formation model (with no disruption but with a
magnitude threshold; Figure~2c) looks much different than the
observed distribution for the Antennae cluster system
(Figure~7).  Instead of having the highest density of clusters per
unit log$\tau$ at older ages, the observations have the highest
density of young clusters. This is demonstrated more clearly in Figure
2 from Fall \etal\/ (2005). They find that a plot of log($dN/d\tau$)
vs. log$\tau$ for the Antennae clusters has a slope that declines
with a value $\approx -1$ out to an age of a few $\times$ 10$^8$
yrs, rather than being flat, as would be the case for constant
cluster formation with no disruption (see Figures 3 and 4).  Hence, if
the Antennae has had a history of roughly constant cluster formation
during this period, clusters must be disrupted at a rate of
$\tau$$^{-1}$.  It is not possible with the current observations to
determine whether the same relationship extends beyond a few $\times$ 10$^8$
yrs for masses $\geq 3\times10^4$~M$_{\odot}$, due to potential
stellar contamination.

The same result is apparent from an examination of Figure~4. Only the
80\% and 90\% [i.e., slopes of $-0.7$ and $-1.0$ in the
log($dN/d\tau$) vs. log$\tau$ diagram] infant mortality laws agree
with the data. None of the other disruption mechanisms comes close to
reproducing the large fraction of observed young clusters.

Perhaps the high density of young clusters is due to a recent burst of
cluster formation in the Antennae? This is likely to be true of
course, since the galaxies are currently merging and regions like the
overlap region have large numbers of very young clusters. However, if
we look at each of the WFPC2 chips independently
(Figure~8), we find that  the log luminosity vs. log$\tau$
diagrams for all the chips look quite similar, with the largest
numbers of clusters always having ages $<$ 10$^7$ years!  Since it is
not possible for the galaxy to synchronize a burst of star formation
over its entire disk (i.e., assuming a ``signal speed'' of $\approx$ 30
km s$^{-1}$ [Whitmore et al. 1999] requires approximately 500 Myr for
one side of the galaxy to communicate with the other side 10 kpc
away), we can only conclude that the vast majority of the clusters are
being disrupted almost as fast as they form, making it appear that the
largest number of clusters is always the youngest.

If one looks closely at Figure~8, or at the corresponding
log($dN/d\tau$) vs. log$\tau$ diagram shown in Figure~9, there
actually are small differences on the various WFPC2 CCDs due to
variations in the local cluster formation histories. For example, the
number of clusters at the youngest ages are especially high for the
overlap region (pixel values $<$ 370 on WF3; i.e., the dusty overlap
region; see Figure 5b from Whitmore \etal\/ 1999), compared to regions
where there appears to be less recent formation, (for example,
pixel values $>$ 370 on WF3, which is the dust-free region containing
several clusters with ages $>$ 100 Myr). However, these are secondary
effects; the primary correlation is the rapid decline in the number of
clusters as a function of age for all of the CCDs. We will return to
this point in \S 3.2.

This behavior is not confined to mergers and starburst galaxies, such as
the Antennae.
In typical spiral galaxies (e.g., M51 - Bastian \etal\/ 2005, Figure
10; M83 - Harris et al. 2002; LMC - Hunter et al. 2003; NGC 6745 - de
Grijs et al. 2003c; M51 and M101 - Chandar et al. 2006; M33 - Chandar
et al. 1999a,b; SMC - Rafelski \& Zaritsky 2005), the observational data
almost universally show that the density of clusters per unit
log$\tau$ is roughly constant.
The fact that such apparently
different galaxies have such similar cluster age distributions is one
more demonstration of the universal nature of cluster demographics,
and highlights the dominant role cluster disruption appears to play  
in all
galaxies.

The phenomenon of star cluster disruption is not surprising.  We are
familiar with the process both observationally (e.g., Wielen 1971;
Bica et al. 2001; Rockosi et al. 2002) and theoretically (e.g., Fall
\& Zhang 2001; Vesperini \& Zepf 2003; Baumgardt \& Makino 2003)).
What is surprising is that {\it
massive, compact} clusters, which might be expected to withstand many
of the disruption mechanisms which are important for low mass,
diffuse clusters (e.g., two-body relaxation), appear to have disruption
time scales that are comparable to those of low mass clusters.

In particular, we note that this is dramatically
different from the canonical picture for open clusters developed by
Wielen (1971), which showed essentially no cluster disruption for the
first $\approx$ 100 Myr (see Wielen 1971, Figure~13).  However, it does match
the recent results from Lada \& Lada (2003) for the young embedded
clusters in the solar neighborhood, since they find that $\sim90$\% of
these clusters are disrupted in the first 10~Myrs.

A natural explanation for this high infant mortality rate is that most
of the YMC's may be unbound when they form (or rapidly become
unbound), and dissolve on the order of a few crossing times (e.g., see
discussion in Fall \etal\/ 2005).  Briefly, the energy and momentum
input from massive, young stars to a protocluster are roughly
proportional to the number of massive stars, and to its mass.
Therefore, such an internal process would be expected to operate
roughly independent of cluster mass, and also independent of host
galaxy properties.

\subsection{Selection Effects and Artifacts}

A number of selection effects and artifacts must be taken into account
before we are able to compare our synthetic data with the
observations.  The first selection effect is a magnitude threshold, as
shown in Figure 2b.
As discussed in \S 2.3, this magnitude threshold, when coupled with
the dimming of clusters with age, can cause serious difficulties when
attempting to determine mass and age distributions over wide dynamic
ranges.

The second effect that we consider is extinction and reddening due to
the presence of dust.  Dust is most closely associated with very young
clusters.  This has been modeled by adopting the Mathis (1990)
coefficients for the various filters, and a time dependence defined by:
$\mbox{A}_V \mbox{(log} \tau) = -2.5\times\mbox{log}(\tau/yr) + 18$
[which results in a mean value of A$_V$ = 3 magnitudes at log$\tau$
= 6.0, $A_V$ = 1.0 magnitudes at log$\tau$ $\geq$ 6.8, and $A_V$ = 0 magnitudes at log$\tau$ $\geq$ 7.2].
Noise is added by
multiplying the A$_V$ value by a random number from 0 to 1 and
then adding a mean value of A$_V$ = 0.2 to represent random foreground
dust.
The time dependent component is consistent with
the extinction distribution $A_V(\mbox{log}\tau)$ found for the
Antennae by Whitmore \& Zhang (2002) and by Mengel \etal\/ (2005).
Using other time-dependent extinction laws (e.g., Charlot \& Fall 2000)
has  only minor effects.

The third artifact we include in the model is the effect of systematic
errors introduced by our age-dating method.
In Figure~7 we show the luminosity versus age diagram for the clusters
in the Antennae galaxies (Whitmore et al. 1999), with ages derived
using the technique described in Fall et al. (2005).  The most
prominent example of an age-dating artifact in this figure is the
apparent gap in the distribution for clusters with ages in the range
log$\tau$ $=7.0\rightarrow7.3$ years
(i.e., $10 - 20$ Myr), which is produced
by the onset of the red supergiant phase.  During this phase, the
integrated cluster colors change so abruptly that the fitted ages, in
the presence of observational errors, become degenerate, with a strong
tendency to avoid values just above $10^7$ yr.
This ``10 Myr artifact'' is found in a wide range of data sets where
age-dating techniques employing broad-band colors are used (e.g., Fall
et al. 2005; Gieles et al. 2005; Harris et al. 2004).

We have modeled biases due to age-dating by simulating clusters with
an even distribution in log$\tau$ and mass, adding noise to the
predicted
photometry, and then running the simulated data through our normal
age-dating technique (which is described in Fall \etal\/
2005). This procedure is sensitive to the prescription used to derive
photometric uncertainties and noise for the simulated cluster
population.  We utilize a procedure similar to that described in
Bastian \etal\/ (2005) for estimating the photometric uncertainty as a
function of magnitude in all filters, based on the measured
photometric uncertainties for the Antennae data.  This assumes that
the uncertainty in the observed magnitude of a cluster, $\Delta m$,
depends on the magnitude, m, as 
$\Delta m=10^{d_1+d_2\times m}$,
where the values of $d_1$ and $d_2$ are determined empirically for
each filter.  In addition, we add scatter to the relationship between
magnitude and photometric uncertainty to better mimic a real dataset.
For the F658N filter, we find that there is larger scatter about
this relation for young clusters which have strong $\mbox{H}\alpha$
emission, and we introduce an additional age-dependent photometric
uncertainty for simulated clusters younger than log$\tau$ = 6.8 yrs
in this filter.  The resulting input age versus output (i.e., fitted) age
is shown in Figure~10, for clusters more massive
than $3\times10^4$~\msun\/ (the lower mass limit adopted in Fall et
al. 2005).  We find that regardless of the mass cut we use (up to
$10^6$ \msun\/), the gap in the age range $10-20$~Myr does not go
away.

Note that this appearance contrasts with Figure~2 in Gieles et
al. (2005), where the gap appears to be present only at lower masses,
but then fills in for higher mass clusters.  We suggest that this is
due to an underestimate in their models of the true photometric
uncertainties at higher masses. As in this work, the Gieles et
al. (2005) simulations use the SED models for both the input and
output.  Hence if the photometric uncertainties are underestimated, the
age dating algorithm will always find the ``right'' answer.  In
reality, there will be a mismatch between the models and the
observations, resulting in a 10 Myr gap at all masses, as seen in the
actual data.

Figure~10 allows one to see in more detail where the data from the
``gap'' is repositioned.  While there is a clear, well-defined
correlation between the input and output ages, the reason for the 10
Myr artifact can be seen as a repositioning of many of the clusters
with ages between log$\tau$ = 7.0 and 7.2 to an age of log$\tau$ =
7.3.

A second prominent feature in Figure~10 is the filled region of points
between log$\tau$ of 6.0 and 6.4 yrs.  This feature implies that
clusters with intrinsic ages between 1 and 2.5 Myr cannot be separated
using integrated colors and narrow band photometry alone.  This is not
surprising, given that the $H\alpha$ emission predicted by models
``saturates'' at these very young ages, hence we see a large
concentration there.

Other potentially important selection effects and artifacts  
are: limited spatial resolution (i.e., confusion between clusters and
stars), blending of objects, and stochasticity for low mass clusters
(e.g., below 10$^4$ \msun\/ the chances of having a single O star
becomes much less than unity, which can affect the photometric colors
dramatically; see e.g., Cervino, Valls-Gabaud, \& Mass-Hesse
2002; Ubeda, Maiz-Apellaniz, Mackenty 2006; Oey \& Clarke 1998). These and other
effects will be simulated in more detail in future versions of the
model.

\section{Results}

Two topics are examined in the current paper in order to exercise the
model described in the previous section.  The first looks
at whether the model can account for the M$_V$(brightest)
vs. log(N) diagram, as shown in Figure 1.  The second topic is the
development of a toy model for the Antennae galaxies. Paper II will
extend the analysis to datasets for a number of other nearby galaxies
(e.g., M101, M51, etc.).

\subsection{A Monte-Carlo Simulation of the M$_V$(brightest) vs. Log (N)
Diagram}

As briefly described in the Introduction,
the M$_V$(brightest) vs. log(N) diagram (Whitmore 2003, originally
presented in 2000 as astro-ph/0012546)
provided much of the initial impetus for the development of the
framework developed in the present paper, since it was consistent with
the idea that mergers, starbursts, and normal spiral galaxies all
share a common universal power law luminosity function $\phi(L)$ for
clusters. Larsen (2002) further developed this basic idea by  
performing Monte
Carlo simulations of cluster populations. This study supports the
conclusion that the magnitude of the brightest cluster is controlled
by statistics rather than by physics. Additionally, Larsen (2002) shows
that the observed spread in the scatter of the relation can also be
explained by statistics.  Several other studies have also begun to
address the important role this ``size-of-sample'' effect can play in
the observed properties of star cluster systems (e.g., Billet \etal\/
2002; Hunter \etal\/ 2003; Weidner, Kroupa \& Larsen 2004).

Figure 1 shows an updated version of the database from
Whitmore (2003). Several
more galaxies have been added, primarily from the
Larsen (2005) compilation, and the values have been updated for
individual galaxies where more recent work has become available.

Here, we present a test of the power law index $\alpha$ over absolute
magnitudes ranging from $M_V$ of $-9$ to $\sim-16.5$.  Figure~11
shows a
comparison between the observational data presented in Figure 1 with
sets of Monte-Carlo simulations of 500 cluster populations, which are
randomly drawn from a power law luminosity function with values of the
index $\alpha$ ranging from $-1.5$ to $-3.0$ and ages between 1 and
100 Myr (i.e., the brightest clusters are essentially always found in
this age range).  The 90 \% infant mortality law is assumed for this
particular simulation.  Several other star formation histories have
also been tested (e.g., constant with time, Gaussian bursts at various
times) with similar results. An observational cutoff brighter than
M$_V=-9$ has been imposed on both the observations and the simulations.

The figure shows that values of $\alpha$ equal to $-1.5$ and $-3.0$
are clearly ruled out. Values of $\alpha$ between $-2.0$ and $-2.4$
all fit the observations to some degree, with values of $-2.0$
matching the slope, but showing an offset of about 1 magnitude, and
values around $-2.4$ showing no offset, but having a flatter slope
than the observations. Values of the fits are included in Table~1. It
is likely that part of the offset is due to many of the datasets not
being complete to M$_V$ = $-9$; hence the value of the number of
clusters in the sample, N, is underestimated. In principle, the slope
$\alpha$ should only reflect the underlying luminosity function.
However, different selection effects (such as the imposed cutoff of
M$_V$ = $-9$), can affect the measured slope.  For this reason, we
consider the $\alpha$ = $-2.2$ fit as the best compromise between
matching the slope and minimizing the offset, and estimate an
uncertainty of 0.2.

Figure~12 shows a comparison between the scatter in the observations
and the simulation for our adopted $\alpha=-2.2$ model. Based on the
similarities between the two, we conclude, as Larsen (2002) did, that
the observational scatter can be explained just by statistics.  There
is no indication that any ``special physics'' is needed to explain the
differences between, for example, the cluster systems of mergers and
spiral galaxies.

We conclude that both the correlation between M$_V$(brightest) and
log(N) and the scatter in the correlation can be explained by
statistics if all star forming galaxies have the same universal
initial luminosity function. The only difference is the normalization,
with mergers having thousands of clusters brighter than $M_V=-9$, and
spirals having tens of such clusters.  One possibility for the much
larger number of clusters in mergers is that the conditions for making
young massive clusters are {\it globally} present while in spiral  
galaxies
they are only {\it locally} present (i.e., in the spiral arms; Whitmore
2003).

\subsection{A Model of the Antennae}

\subsubsection{Background}

Because of its proximity (19.2~Mpc), large collection of young massive
clusters, and extensive set of multi-wavelength observations (e.g.,
Zhang, Fall \& Whitmore 2001), the prototypical merger NGC 4038/4039
(the ``Antennae'') currently represents the best dataset for testing
the model described in this paper.  Here, we briefly summarize the
observations and analysis from these previous works.

The {\it HST} observations of the Antennae galaxies are described
more fully by Whitmore et al. (1999).  The images were taken in 1996
January by the Wide-Field Planetary Camera 2 (WFPC2) with each of the
broadband filters F336W ($U$), F439W ($B$), F555W ($V$), and F814W
($I$), and with the narrowband F658N ($H\alpha$) filter. Approximately
14,000 point-like objects (stars and clusters) were detected.

We estimate both the age ($\tau$) and extinction ($A_V$) of each
cluster by comparing the observed magnitudes in the five bands with
those from stellar population models, as described in Fall et
al. (2005).  More specifically, we use the Bruzual \& Charlot (2003)
models with solar metallicity and Salpeter IMF.
While this metallicity is appropriate for comparison with young
stellar populations ($\lea1$~Gyr), estimates of ages for older stellar
populations should be viewed with caution, as they suffer from the
well-known age-metallicity degeneracy, and likely have abundances
lower than the solar value. Because dust is prevalent and patchy,
it is important to estimate the internal effects of dust individually
for each cluster.  However, for compact star clusters in dusty regions
it is not obvious that either a simple foreground screen
or an attenuation law which assumes gas and dust mixed with stars
(e.g., Calzetti, Kinney \& Storchi-Bergmann 1994) is appropriate.  We
therefore experimented with both a Galactic-type extinction law
(Fitzpatrick 1999) and the absorption curves determined from starburst
galaxies by Calzetti et al. (1994) for the age dating procedure.  The
main difference between the two is the value of $R_v$, which is 3.1 for
Galactic extinction laws and 4.05 for the Calzetti absorption law.
We found relatively minor differences in the results,
with little change in the derived cluster ages, and masses
which were $10-20$\% higher for clusters younger than $10^7$~yrs when
the Calzetti law was assumed.

As briefly described in the Introduction, our earlier analysis of the
Antennae dataset provided some of the motivation for the current paper.
Of particular relevance for the current work, Fall (2006) notes that
the mass and age distributions over limited ranges for the Antennae
cluster system are {\it independent} of one another.  This means that
the joint distribution $g(M,\tau)$ must be the product of the mass and
age distributions, $\psi(M)$ and $\chi(\tau)$, which roughly follow
the relation:

\begin{equation}
g(M,\tau) \propto M^{\beta}\tau^{\gamma} \propto M^{-2} \tau^{-1}~~~~~
\mbox{for}~~ \tau \lea 10^{7}(M/10^{4} M_{\odot})^{1.3}~ \mbox{yr}
\end{equation}

\noindent as used in \S2 to define the approximate mass$-$age
distributions of the massive young clusters.

In general the luminosity function is {\it not} independent of the
mass and age distributions. However, under certain circumstances
(e.g.,  when the mass distribution is a power
law and is independent of the age distribution)
the luminosity and mass functions can both be power laws with the
same exponent.
This appears to be the case for the Antennae, with
$\phi(L)dL \propto L^{\alpha} dL$; and $\alpha\approx-2$ for
$-14<M_V<-6$
(Whitmore \& Schweizer 1995; Whitmore \etal\/ 1999),
and $\psi(M)dM \propto M^{\beta}dM$, with $\beta \approx -2$ for
$10^4-10^6~M_{\odot}$ (e.g., Zhang \& Fall 1999; Fall, Chandar, \&
Whitmore 2006).

Our purpose here is not to revise the previous work, but rather to
extrapolate to both higher and lower mass ranges.  Applying these
relations,
the model allows us to take the next step and move beyond the
observed cluster system, and compare predictions with the
integrated properties of a galaxy (e.g., the total stellar mass and
total luminosity).
For example, to the degree that our models fit the integrated galaxy
properties,
we can have some confidence that extrapolations outside of the
original age and mass ranges (as specified in equation 1) are
reasonable.

In addition, we can test a variety of the other assumptions that go
into the model. Does our assumption that all stars form in
clusters, and that the field stars are the remnants of disrupted
clusters, give values for the luminosities of the clusters and field
which are consistent with observation?  
Can we predict other distributions (or different
projections of the data, e.g., color-color and color-magnitude
diagrams) which were not used in the original determination of
$\alpha$, $\beta$, and $\gamma$ above?  Does an extrapolation to the
highest observed mass support the recent suggestion that the Antennae
system has a fixed upper cluster mass?  While there is already some
evidence that parts of the model are consistent with observations for
a few other nearby galaxies (see, for example, \S 1 and Figures 1, 11,
12, 15), in Paper 2 we will extend the analysis to a larger sample to
better test the universality of the basic framework.

\subsubsection{4-Component Model of the Antennae}

We begin by constructing a model for the Antennae which assumes four
cluster formation components, which are derived from previous
information (Whitmore \etal\/ 1999). The four components are as
follows: Component \# 1 - An initial Gaussian burst with an age of
13~Gyr and $\sigma$ of 1 Gyr that formed the population of old
globular clusters.  Component \# 2 - A constant rate of cluster
formation over the past 12 Gyr similar to the cluster formation rate
in typical spiral galaxies.
Component \# 3 - A Gaussian
burst of cluster formation $500\pm 100$~Myr ago that was triggered when
the galaxies first encountered each other and the long tidal tails
were ejected.  Component \# 4 - An increased rate of cluster formation
during the past 100 Myr responsible for most of the recent
star/cluster formation.  The last two components are also motivated
by the simulations of the dynamical interaction between the two
galaxies (Barnes 1988), which suggested that the galaxies first
encountered each other a few 100 Myr ago, and are now in their
second encounter.
{\it Hence, the 4-component model
essentially has no free parameters}. It is
an attempt to construct a model using previous determinations
of $\beta$, $\gamma$, and the cluster formation history of the
Antennae, and an assumption that the power laws continue beyond
the ranges where they have been observed directly.

The observations and derived properties of the Antennae are shown in
the left column of panels in Figure~13.
In order to restrict the Antennae sample to actual star clusters and
not individual stars, we make a cut in absolute luminosity at $M_V=-9$
(see Whitmore \etal\/ 1999).  Following Fall et al. (2005), we also
impose a mass limit ($\geq 3\times 10^4~M_{\odot}$).  Note that in the
age versus mass figure of the Antennae clusters this mass limit begins
to show incompleteness for clusters starting at ages $\sim10^8$ years.

The model uses a value of $\beta = -2.0$ and the two-stage
disruption law with 90 \% infant mortality, as motivated above.  This
4-component Antennae model is shown in the middle column of Figure
13.
We remind readers that the apparent gap in the age distribution around
10 Myr (i.e., Figures 7 and 10) is due to artifacts from the
age-dating algorithm. While our modeling of this artifact (\S2.5) is
able to subjectively reproduce the major features, some of the
small-scale remaining differences (e.g., the narrowness of the 10 Myr
artifact) are still not very well represented.

There is reasonable agreement between the 4-component model
and many of the diagnostics and observations (e.g., the log($dN/d\tau$)
vs. log$\tau$ diagram) right out of the box (i.e., using
previous determinations of $\beta$, $\gamma$, and the cluster  
formation history
of the
Antennae).  However, a closer look reveals three issues:
1) the lack of intermediate-age clusters around $100 - 300$~Myr
(besides being evident in the luminosity-age and mass-age diagram,
this deficit can also be seen in the color-color diagram as the slight
misplacement of the small enhancement around $U-B$ = 0.0 and $V-I$ =
0.5 mag; see also Figure 17 from Whitmore \etal\/ 1999); 2) a slight
deficit in the number of young clusters with ages 1 $-$ 10 Myr; and 3)
a slight enhancement in the predicted number of very luminous
and/or very massive clusters.

The first difference can be fixed if we move the 500~Myr
burst to an age of about 200 Myr. There is independent
evidence for this in the numerical modeling of the Antennae
(Barnes 1988).
The second difference is the need for an enhancement
at very young ages (i.e., 1 $-$ 10 Myr). This is probably related to the
enhancement we see in the overlap region (on WF3 as already noted in
\S2.4).
The third difference suggests either the possibility of an upper mass  
cutoff
for the clusters, or
a slightly steeper initial mass function (i.e., $\beta$ = $-2.1$
rather than $-2.0$; these possibilities are explored in \S 3.3).

\subsubsection{5-Component Model of the Antennae}

Our second attempt at making a model takes into account the
shortcomings of the 4-component model mentioned above.  Because of the
addition of a $1-10$~Myr component (to address the second shortcoming), 
we now refer to this as the
5-component model. In addition, we relax the constraint that only the
canonical values of $\gamma=-1$ (i.e., the coefficient describing the
infant mortality law) and $\beta=-2.0$ must be used.

The 4-component model had a total galaxy luminosity that is
considerably too high (i.e., $M_{V,galaxy}=-23.0$, compared with the
estimated $M_{V,galaxy}=-21.7$ for the Antennae; Whitmore \etal\/
1999).  We therefore used an 80 \% infant mortality law (i.e., $ 
\gamma=-0.7$)
for all
components of our 5-component Antennae model which results in a
smaller total galaxy luminosity and mass. This is because the predicted
cluster population is normalized to the observed Antennae cluster
system, so that the lower rate of cluster disruption implies that fewer
clusters were formed initially.  This also creates a smaller number
of very young clusters, which is accounted for by increasing the
$1-10$~Myr component slightly.
From Figure~4 it appears that the 80\% and 90\% infant mortality
laws both give reasonable fits.
Finally, we change the power law mass index $\beta$, from $-2.0$ to
$-2.1$, as justified in \S3.3.  These two changes reduce the total
galaxy mass to $\sim4 \times 10^{10}$ \msun\/ and total M$_{V,galaxy}$
to $-22.2$ (much closer to the observed value of $-21.7$; see
Table~2). The results from the final 5-component model are shown in
the right column of Figure~13.

As an additional consistency check, we
compare the predicted UV luminosity in clusters
from our 5-component model with the observed luminosity.
The model predicts that $\sim7$\% of the UV
luminosity should come from clusters.  This is quite similar to
the value of $\sim$ 9\% measured for the Antennae cluster system by
Whitmore \& Zhang (2002).

While we could have continued to tweak various parameters in our
model to exactly match the observations of the Antennae, we do not
think this approach is useful, since there are a large number of
potential variables, many of them coupled. Instead, our approach has
been to only make changes where there is an independent indication
that a value should be changed (e.g., the 500 Myr component should be
changed to 200 Myr; see previous section).  The only exceptions are
the variables $\beta$ and $\gamma$, which we allowed to vary (within
the uncertainties) in order to address the discrepancy between the
predicted and observed total galaxy mass and luminosity.

\subsection{Is There an Upper Mass Cutoff for Clusters in the Antennae?}

Is there a physical limit with which star clusters in the Antennae can
form, or does the most massive cluster just result from statistics?
Gieles et al. (2006) have suggested that the former is true in the
Antennae.  In \S2.2 and \S3.1, we showed how the maximum cluster mass
increases with the total number of clusters.  Given
its very rich population of clusters, the Antennae
is an ideal galaxy to test whether there is any evidence for a
physical upper mass cutoff with which clusters can form.

As briefly noted in \S 3.1, the cluster luminosity function for  
the
4-component Antennae model (shown in the middle column in Figure~13)
agrees quite well with the data for the fainter magnitudes, but the
simulations slightly overpredict the number of the brightest clusters.
This results in an observed luminosity function which is too flat,
especially at the bright end.

To explore the possibility of a cutoff to the cluster mass function,
we started with a slope $\beta\sim2.0$ model, which is known to hold for
young $10^4-10^6~M_{\odot}$ clusters in the Antennae (Fall, Chandar,
\& Whitmore 2006).
Because the high mass end of the cluster
mass function suffers from small number statistics
the total number of clusters, N,
used for each realization was determined empirically (by matching the
simulation to the observed number of Antennae clusters with masses
between $5\times10^4~M_{\odot}$ and $5\times10^5~M_{\odot}$).  The number
of predicted clusters with masses $>10^6~M_{\odot}$ was then compared
with the observed masses of clusters in the Antennae.

Figure~14 shows the result from our Monte-Carlo simulations.  The left
column shows a comparison of the Antennae values (dashed line) for the
brightest, 3rd brightest, and 10th brightest cluster with similar
ranked clusters from the simulations for $\beta=-2.0$.  There does
appear to be a difference between the simulations and data, with the
simulations predicting somewhat larger masses (i.e., at the 1 $-$ 2
$\sigma$ level) for all three panels.  However, we note from Figure~14
that this might also be explained by a small difference in the value
of $\beta$.  Simulations showing the results assuming $\beta=-2.1$ and
$\beta=-2.2$ are shown in the middle and right panels respectively.

Figure~14 shows an asymmetric tail toward larger masses for all tested
values of $\beta$,
predicting that in a few cases, cluster masses as high as 10$^8$
\msun\/ or even (very rarely) 10$^9$ \msun\/ might be expected if
there is no upper mass cutoff. This high mass tail might explain the
existence of W3 in NGC 7252, which has a dynamical mass $8\pm2
\times10^7$~\msun\/ (Maraston et al. 2004).  

For the Antennae, our
sample contains a number of clusters more massive than the proposed
limit of
$1.9\pm0.6 \times10^6$~\msun\/ suggested by Gieles et al. (2006).
In fact the $\beta=-2.1$ simulation in particular provides a good
match to observed cluster masses $\geq10^6~M_{\odot}$ in the Antennae,
and is within the observational uncertainties for the mass function
measured for young clusters in this galaxy.
Hence, we find that the cluster mass distribution in the Antennae
galaxies {\it does not require} a truncation at the high mass end, and
can be explained by sampling statistics and an initial mass function
slope $\beta\sim-2.1$.  Note however, that the small number of very
massive clusters in our sample results in large uncertainties at this
end of the cluster mass function.  Therefore the current observations
cannot exclude an actual cutoff in the Antennae mass function above
$\sim\mbox{few}\times10^7~M_{\odot}$.

\section{Discussion}

\subsection{Potential Anomalies and Predictions}

We have demonstrated that our framework and simple toy model can
explain the basic demographics of star clusters in the Antennae
galaxies, and in Paper 2 we will show that the same framework can be
used to successfully model other nearby galaxies.  However, it
is possible to think of apparent exceptions to the
framework outlined above.

One example of a potential anomaly, the very bright cluster in
NGC 1569, has already been discussed extensively in the literature
(e.g., Whitmore 2003; Billet \etal\/ 2002; Larsen 2002).  We first
note that this cluster is no longer as discrepant as previously
believed, since the number of detected clusters in the sample of
Hunter \etal\/ (2000) ($N = 18$ with M$_V$ $<$ --9) is larger than in
the original study by de Marchi \etal\/ (1997) ($N = 7$). In \S 3.1 we
found that the scatter predicted by statistics is consistent with the
observed scatter. The position of NGC 1569 in Figure 1 using the new
Hunter data is $\lea2\sigma$ (predicted scatter) from the mean
relation. Hence the point is no longer very discrepant.

We should, however, keep in mind that subtle selection effects can
alter the results for individual galaxies. For example, Billett
\etal\/ (2002) and Whitmore (2004) both point out that in a sample of
star forming dwarf galaxies, the ones with the brightest
clusters (i.e. the 1 $-$ 2 $\sigma$ high points) are more likely to
have drawn attention to themselves and to have been observed than
dwarf galaxies that have not formed bright clusters (i.e., the 1 $-$ 2
$\sigma$ low points).

The second apparent anomaly we discuss is the lack of intermediate-age
globular clusters in the Milky Way. If the majority of stars form in
clusters, and the initial cluster mass function is universal, then the
fact that there are many intermediate-age stars (for example the Sun
!), implies that there should be some intermediate-age (i.e., a few
billion year old) globular clusters.

To some extent this may just be a matter of semantics.  The tail of
the distribution of old ``open'' clusters may simply be
intermediate-age globular clusters, as suggested in Whitmore
(2003). While there are only a handful of known old open clusters, we
should keep in mind that the catalogs of open clusters are only
complete within a distance of about 1 $-$ 2 kpc of the sun; hence the
volume for the entire Milky Way is roughly 100 times larger.  There
are certain to be many more clusters in this category hidden behind
the dust in the plane of the Milky Way. Indeed, in neighboring
spiral galaxies, where we are able to see the entire disk, intermediate-age
globular clusters are now being found (e.g., in M31 see Puzia, Perrett, \&
Bridges 2005; Beasley \etal\/ 2004; Burstein \etal\/ 2004; Li \&
Burstein 2003; Barmby \& Huchra 2000; but see Cohen et al. 2005 concerning
the inclusion of asterisms in some of these compilations;
in M33 see e.g., Chandar et al. 2002, 2006). 

We might also note that due to a combination of the various
disruption mechanisms and the completeness limit, the number of
observable intermediate-age clusters is expected to be relatively
low. For example, in our 5-component simulation of the Antennae, only
$\sim10$ clusters are predicted in the range 1 $-$ 10 Gyr with M$_V$ $>$
$-9$, and only 8 clusters are observed in this region.  The excess of
clusters with ages $\approx$ 13 Gyr (i.e., old globular clusters) in
the model and in the data is due to a strong initial burst, which is
expected to produce roughly the same number of clusters
as are produced in the next $\approx$ 10 Gyr.

The third ``anomaly'' that we address is the apparent difference
between the metallicity distribution of Galactic field stars, with a
broad peak around [Fe/H] $\approx$ --0.8, and the metallicity
distribution of the globular clusters, which is bimodal with peaks
around [Fe/H] $\approx$ $-1.3$ and $-0.75$ for the Galaxy (see
Figure~8 in Cote 1999).
This is highlighted in Harris (2003) for M31, and by Beasley \etal\/
(2003) for NGC~5128.  In the simplest version of our model,
we would predict that the metallicity distributions of the field stars
and the clusters would be identical.


Harris (2003) points out that another way of thinking about this is
that the specific frequency (number of clusters per galaxy luminosity)
needs to be roughly 5 times larger for the blue (metal-poor) clusters
than the red (metal-rich) clusters. He argues that the efficiency of
{\it
formation} of the blue clusters must somehow be higher. However,
another possibility is that the difference lies in {\it
disruption} rather than formation mechanisms.  In the
context of the current paper, this might simply be explained by
changing the infant mortality rate from 90 \% for the current epoch to
80 \% during the epoch when the old (blue) globular cluster population
formed. This would result in $\sim4$ times more blue clusters
surviving infant mortality in the first 100~Myr.  Hence, a very small
change in the survival rate for the blue, metal-poor clusters would be
enough to change the value of S$_N$ by a factor of $\approx$5 without
significantly changing the overall framework of the model.  We refer
the reader to Harris \etal\/ (2006) for a discussion of what
might cause the blue, metal-poor globular clusters to have a lower
disruption rate.  More generally, any correlated variable
(for e.g., orbital motions), which correlates with metallicity and
also couples to disruption rates would achieve a similar result (see
Fall \& Rees 1985 for a discussion of this point).  This topic will be
addressed more carefully in Paper II.

\subsection{Choice of the Initial Cluster Mass Function}

In this paper, we have assumed a single power law for the initial
mass function, since many works point to this form (see references in
the Introduction).  However, recently there have been suggestions that
the initial mass function may be a broken power law (e.g., Mengel et
al. 2005; Gieles et al. 2006) or even a Gaussian (e.g., Fritze-van
Alvensleben 2004; de Grijs, Parmentier, \& Lamers 2005).  A full
treatment of these forms is beyond the scope of this work, particularly
since in our judgement, there are a number of reasons to remain
suspicious of these suggestions.

Mengel et al. (2005) present a cluster mass function for the Antennae
which is best fit by two power laws or a ``broken'' power law (i.e.,
the mass function is steeper at the high mass end, and flattens
towards lower masses).  They use a K-band limited cluster sample,  
which is
much shallower than the $HST$ data referred to here.
As they make clear, their sample selection has very complicated and
poorly understood selection effects, making it difficult to assess
incompleteness, and likely affecting the low mass end of their mass
function.  Given these difficulties, Mengel et al. (2005) caution
against ``over-interpreting'' the presentation of their mass function.
A comparison of the Antennae mass function discussed in \S3.3 and that
presented in Mengel et al. (2005) shows good agreement in the slope at
the high mass end (see Fall \etal\/ 2006).  Their mass function begins
to deviate (becomes shallower) from ours at lower masses, exactly as
expected from a shallower data set without robust completeness
corrections.

De Grijs, Parmentier, \& Lamers (2005) claim that the mass function
for the 1~Gyr cluster population in M82 was initially a Gaussian, and
that there has been essentially no disruption of low mass clusters in
this galaxy. 
They find that the mean mass they derive for the 1 Gyr
population is $\approx 10^5$ \msun\/, essentially the same as a normal
population of old globular clusters. Hence, they argue, there is no
room for the disruption of low mass clusters if this population is to
evolve into a classic globular cluster system in $\sim10$~Gyr.
However, we note that this conclusion is not consistent with their own
data, since all nine of the clusters with ages less than 100 Myr in
Figure 3 of de Grijs et al. (2003a) have masses less than 10$^5$
\msun\/.
Given the high and variable
extinction in this galaxy, we suspect that completeness and
extinction effects may be underestimated for their 1 Gyr population,
resulting in an overestimate of the mean mass of the population.

We conclude by noting that in the Antennae at least, clusters
with masses less
than $\approx 10^8$ \msun\/ and ages less than 100 Myr clearly have a
power law mass distribution down to the limit where clusters can be
separated from stars based on luminosity (Zhang \& Fall 1999; Fall
2004; Fall et al. 2006), hence justifying our assumption of an initial
power law mass function in our model.

\subsection{The Antennae as a Scaled-Up Version of the Milky Way}

Lada \& Lada (2003) make several of the same points about the need for
rapid cluster disruption for very young clusters in the Milky Way,
that are made for the Antennae in this paper (\S 3.2).  In fact, it is
possible to view the cluster demographics in the Antennae as a
scaled-up version of the Milky Way. According to Lada \& Lada (2003),
the Milky Way has roughly 100 known embedded young clusters within a
distance of $\approx$ 1 kpc. The most massive clusters in their sample
are slightly more massive than 10$^3$ \msun.  If we were able to see
the entire disk of the Milky Way the sample would be roughly 100 times
larger. The current star formation rate per unit area in the Antennae
is also roughly a factor of 10 higher than the Milky Way (Zhang, Fall,
\& Whitmore 2003), hence the total enhancement for the number of young
clusters in the sample might be roughly a factor of 1000. Assuming a
universal power law with index $\approx$ --2 for the initial mass  
function, the
most massive clusters with comparable ages to the Lada \& Lada sample
(i.e., $\approx$ 10 Myr) would be predicted to be slightly more
massive than 10$^6$ \msun, just as they are in the Antennae.

Lamers \etal\/ (2005) have used the new catalog of Galactic open
clusters from Kharchenko et al. (2005) with the predictions of the
BL03 cluster disruption model. They find an apparent break in the
$dN/d\tau$ diagram around log$\tau$ = 8.5 yrs, and interpret this
in the context of their model (see \S 2.3 for a discussion) as
evidence for a value of $t_{4}^{dis}$ = 1.3 $\pm$ 0.5 Gyr.  However,
if we add the embedded clusters from Lada \& Lada (2003) to their
sample, most or all of the evidence for a bend in the $dN/d\tau$
diagram goes away.  Hence, the Lamers et al. (2005) result for the
Galactic open cluster $t_{4}^{dis}$ 
may be another example
where selection effects and sample incompleteness (i.e., not including
the young embedded clusters from Lada \& Lada 2003) can produce an
apparent bend in the age distribution, which can be misinterpreted in
the BL03 models as evidence for a specific value of the ``disruption
time''.

In Figure~15 we plot cluster age distributions for the Lada \& Lada
(2003) embedded cluster sample, the Kharchenko et al.  (2005) open
cluster sample, and our Antennae clusters more massive than
$10^5~M_{\odot}$.  In addition, we show the age distribution for
clusters in the SMC using data from Rafelski \& Zaritsky (2005), but
reinterpreted by Chandar, Fall \& Whitmore (2006).
We note that a $dN/d\tau \propto \tau^{-1}$ dependence again fits the
data, although Rafelski \& Zaritsky (2005) quote a power law fit to
this distribution of $dN/d\tau \propto \tau^{-2}$ in their paper 
(see discussion in Chandar, Fall \& Whitmore 2006).

To summarize, Figure~15 shows that the cluster age distribution in
these three very different environments (and covering different mass
ranges) all exhibit the same characteristic behavior, with a
$\sim\tau^{-1}$ decline in the number of clusters with age over the
first $\sim10^9$ years.

\section{SUMMARY}

Motivated by our previous results using age and mass
distributions for the Antennae cluster system, and similar results for
a few other galaxies, we have developed a framework for understanding
the demographics of star clusters, along with a toy model.  This model
incorporates a universal initial power law mass function, selected
formation histories, selected disruption mechanisms, and a
convolution with artifacts and selection effects.
The three key parameters in the models are the power law index for
the mass function ($\beta \approx -2$; primarily affects the
cluster mass distribution at the high mass end and total galaxy mass
and luminosity), the percentage of clusters disrupted in the infant
mortality disruption law ($\approx$ 90 \% per decade of $\tau$;
i.e., $\tau^{-1.0}$ dependencies;
primarily affects the total galaxy mass), and the mass loss
rate $\mu_{ev}$ (primarily affects the shape of the observed cluster
mass
function and the total mass).

A wide variety of
observations can be explained by this simple model. In this particular
contribution we concentrate on the M$_V$(brightest) vs. log(N)
relationship and on extending the range of comparison for observations of
the Antennae galaxies. In Paper II we consider several other nearby
galaxies.

1. The correlation between the brightest young cluster in a galaxy and
the total number of young clusters [i.e., M$_V$(brightest) vs. log(N)]
can be understood as a statistical size-of-sample effect rather than a
difference in the physical process responsible for  the
formation of the clusters. One possibility for the much larger number
of clusters in mergers is that conditions for making young massive
clusters are globally present while in spiral galaxies they are only
locally present (i.e., in the spiral arms).  The diagram is quite
sensitive to the value of $\alpha$ for the initial power law
luminosity function, with relatively good agreement in the range $-2.0
< \alpha < -2.4$.

2. A detailed comparison is made between different cluster disruption
laws and it is demonstrated that the apparent break in the $dN/d\tau$
vs. log$\tau$ diagram used to determine the parameters in the
Boutloukos \& Lamers (2003) model may be produced by  
incompleteness at
the breakpoint.

3. A 4-component model of the Antennae galaxies was first developed
using only pre-existing information
(i.e., 90 \% infant mortality rate,
$\beta$ = $-$2.0, and cluster formation histories from Whitmore
\etal\/ 1999).  The model showed reasonably good agreement with the
data, although there were three minor areas of disagreement. In
addition, the total magnitude predicted for the galaxy by
extrapolating $\beta$ to the low stellar mass regime 
was $\approx$1 magnitude brighter than the Antennae.

These shortcomings were addressed in a 5-component model
of the Antennae galaxies
which agrees with the observations quite well, based on matches to the
magnitude-age, number-magnitude (i.e., luminosity function),
color-magnitude, color-color, mass-age and number-age (i.e., age
distribution) diagrams.  The model employs a two-stage cluster
disruption law with 80\% infant mortality in the first 100 Myr (i.e.,
a $\gamma \approx -0.7$ dependence; similar to the observed value of
$\gamma \approx -1$; Fall \etal\/ 2005), and a value of $\beta = -2.1$
for the index of the initial mass function (within the uncertainties
of previous studies, which were determined directly over a more
limited age-mass domain;
Zhang \& Fall 1999; Fall 2004; Fall \etal\/ 2006). This results in a
predicted total galaxy mass and luminosity, and fraction of UV light
in clusters, that are in reasonably good agreement with the
observations (see Table 2).

4. We find no evidence for a truncation of the cluster mass function
at the high mass end.  In fact, $\beta\sim-2.1$ gives a good match to
the observed cluster mass function above the stellar
contamination limit (see Fall \etal\/ 2006 for a more detailed
treatment).

5. Four potential anomalies are examined and found not to be serious
problems for the framework. The brightest cluster in NGC 1569 falls
well within the statistical uncertainty expected by the
model. Intermediate-age globular clusters have recently been found in
nearby spiral galaxies such as M33 (e.g., Chandar et al. 2002; 2006)
and M31 (e.g., Barmby \& Huchra 2000; Li \& Burstein 2003; and Puzia
\etal\/ 2005).  The difference in the metallicity [Fe/H] distribution
functions of stars and globular clusters in various galaxies may be
due to less effective disruption of the metal-poor clusters (e.g.,
Fall \& Rees 1985; Harris \etal\/ 2006). The claim by de Grijs \etal\/
(2005) that the initial mass function in M82 is a Gaussian is not
consistent with their own data for the young clusters, and is clearly
not the case in the Antennae (Fall \etal\/ 2006).

6. It is demonstrated that the basic demographics of the star clusters
in the Antennae galaxies are consistent with a ``scaled-up'' version
of the local neighborhood of the Milky Way. The required scaling
factor ($\approx$ 1000) is due to the larger volume and higher star
formation rate in the Antennae. This provides general support for the
universal nature of the framework outlined in this paper.

\acknowledgements We thank Francois Schweizer and the referee for comments
which improved the paper.  This work was partially
supported by NASA grants GO-09720-A and GO-10188-A.  $HST$ data are obtained
at STScI, which is operated by AURA, Inc., under NASA contract NAS5-26555.

\clearpage

\begin{figure}
\epsscale{.75}
\plotone{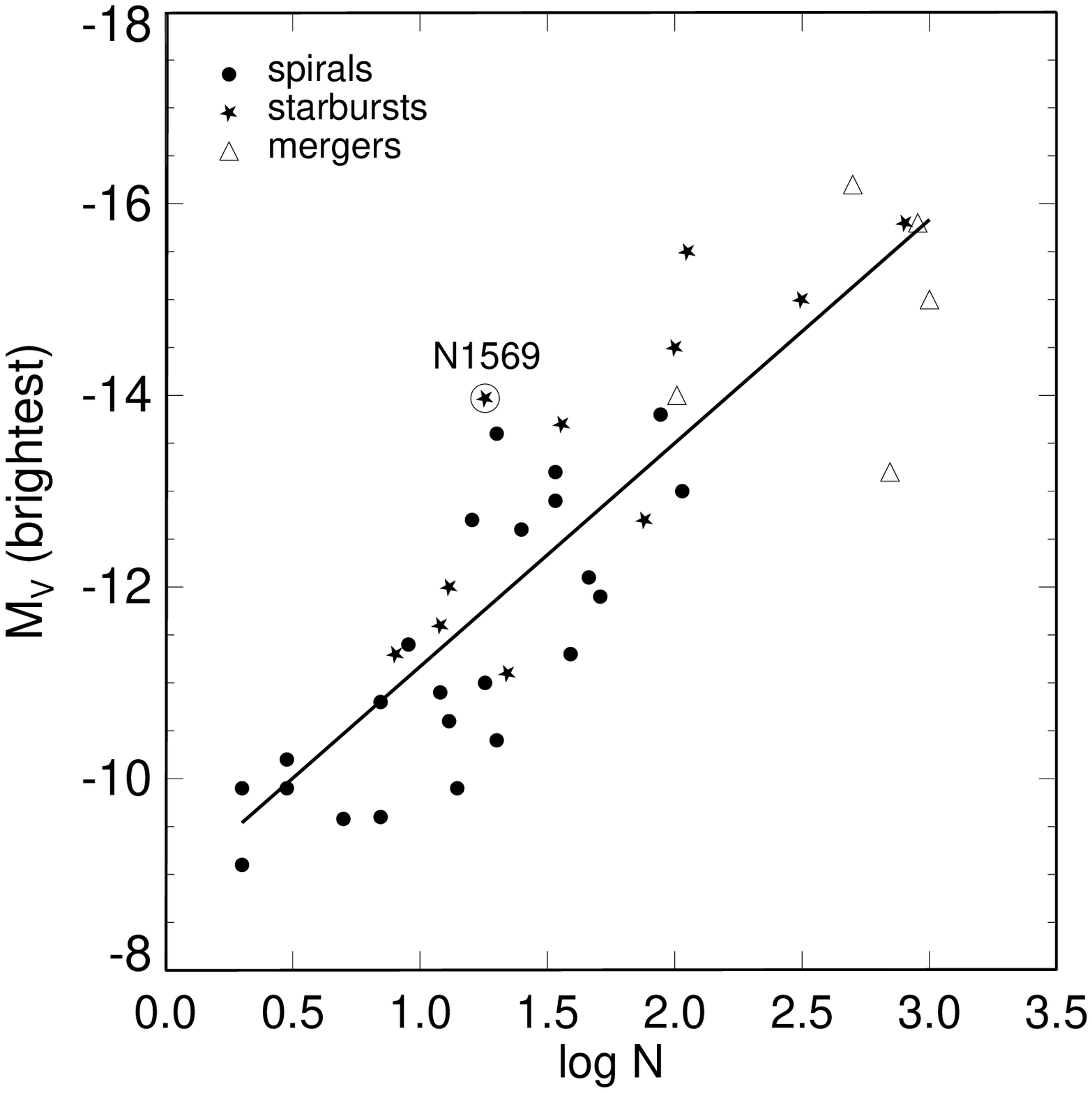}
\caption{Magnitude of the brightest cluster against log
of the number of clusters brighter than $M_V=-9$ for 40 galaxies.  The
solid line shows the best fit, which has a slope $-2.3\pm0.2$.
\label{fig1}}
\end{figure}


\begin{figure}
\epsscale{.75}
\plotone{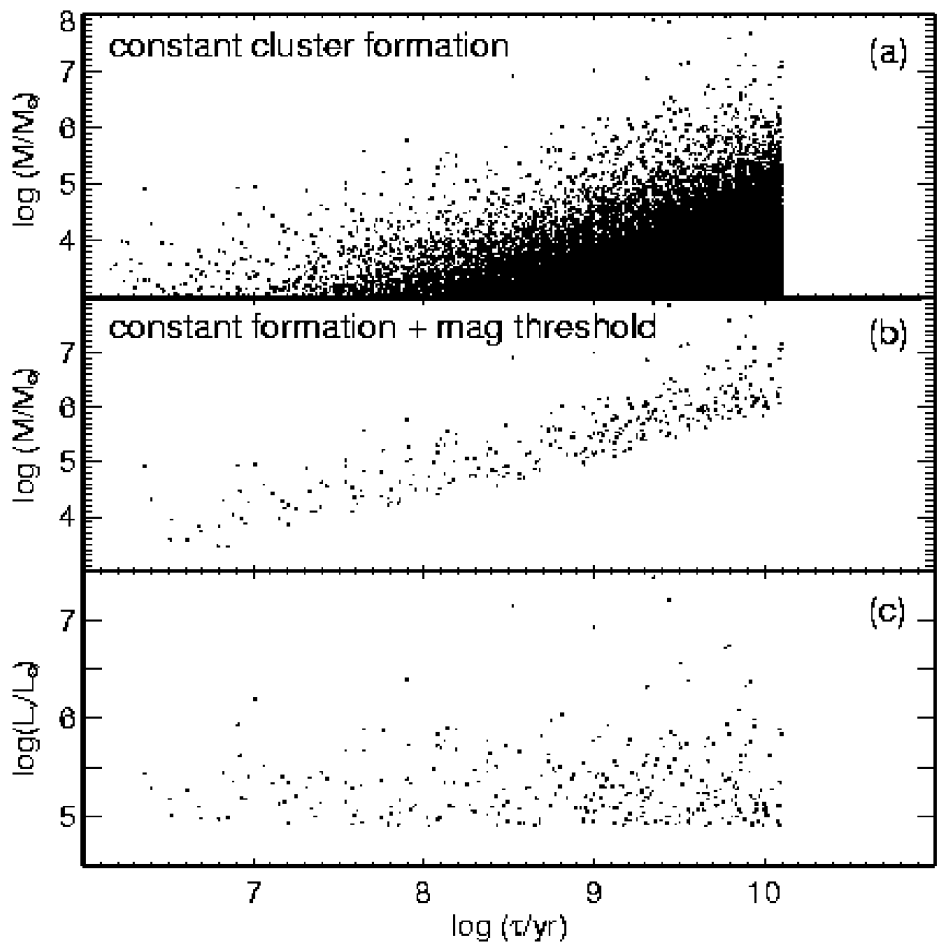}
\caption{Synthetic cluster
population formed at a constant rate for clusters with mass
$> 10^3$ \msun\/.  The top panel shows the age
versus mass distribution, which has been drawn
randomly from a power law mass function which has an index $\beta=-2$.
The middle panel shows the age versus mass diagram for the same model,
but with a V magnitude limit imposed to
mimic observations.  The bottom panel shows the V band
luminosity versus the age distribution of the synthetic cluster
population, including the magnitude limit.
\label{fig2}}
\end{figure}

\begin{figure}
\plotone{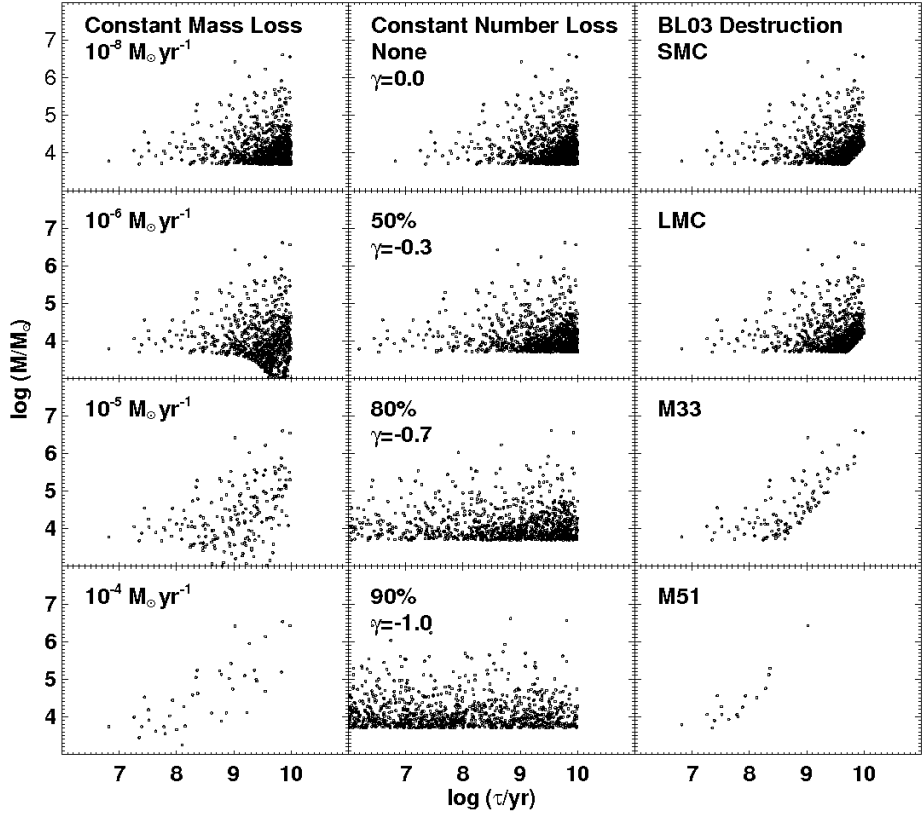}
\caption{Age versus mass for a constant cluster
formation model, which has been modified by twelve different
disruption laws (described in the text).  The left column of panels
shows different values of constant mass loss (i.e.,
two-body relaxation; see Fall \& Zhang 2001), the second column
shows constant number loss (i.e., ``infant mortality''), and the last column
shows the resulting cluster population for the 
mass-dependent disruption model proposed by Boutloukos \& Lamers (2003; BL03).
The parameters for the BL03 disruption law are derived for the LMC by
de Grijs \& Anders (2005), and for the SMC, M33, and M51 by LGPZ05.
Note that for illustrative purposes, we have plotted the infant  
mortality
law
(second column) beyond the first 100 Myr, although we do not
believe it is physically relevant beyond this age (see text).
\label{modeldestruct}}
\end{figure}

\begin{figure}
\plotone{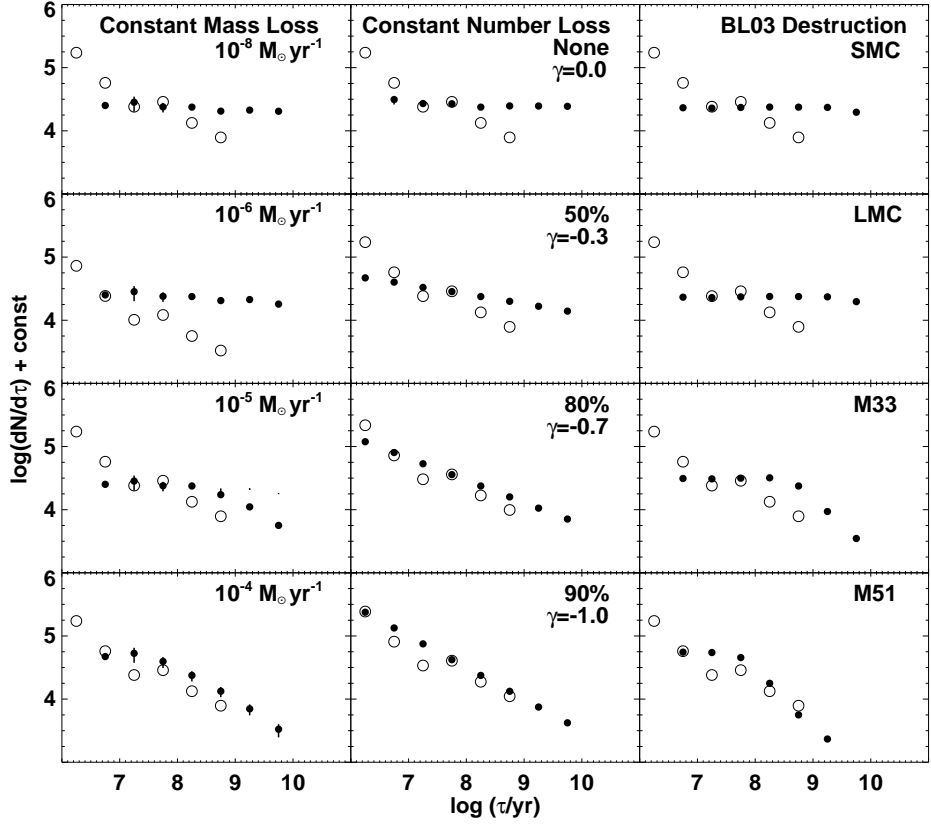}
\caption{Age distribution (with arbitrary offset) for the twelve
simulated models shown in Figure~3 (solid
circles).  For comparison, we show the completeness corrected age
distribution for clusters in the Antennae galaxies with masses
$\geq10^5~M_{\odot}$ (open circles).  
\label{modeldndt}}
\end{figure}

\begin{figure}
\plotone{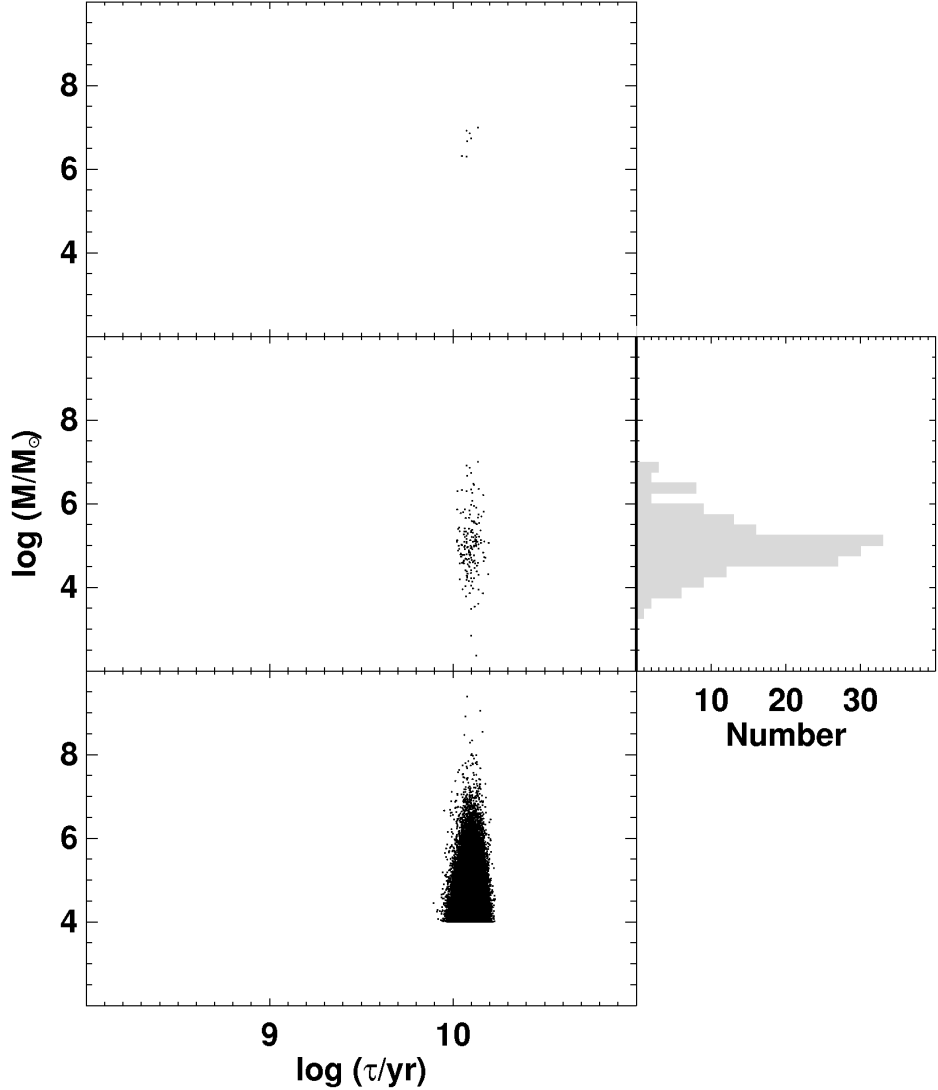}
\caption{Simulation showing original (bottom panel, N$=$200,000),
surviving (middle
panel, N$=$175), and observable clusters (top panel, N$=$7) 
for an ancient Gaussian burst model at the distance
of the Antennae.
These were formed assuming a power law mass function with $\beta=-2$,
a Gaussian burst with a mean age of 12.6~Gyr, and a width of
1~Gyr.  The initial cluster population was then subjected to our
two-stage disruption model, as described in the text.  The grey
histogram shows that the resulting mass distribution after
$\sim13$~Gyr has a peak mass
$\sim10^5~M_{\odot}$, similar to what is found for the Galactic
globular cluster system.
\label{fig:gc}}
\end{figure}

\begin{figure}
\plotone{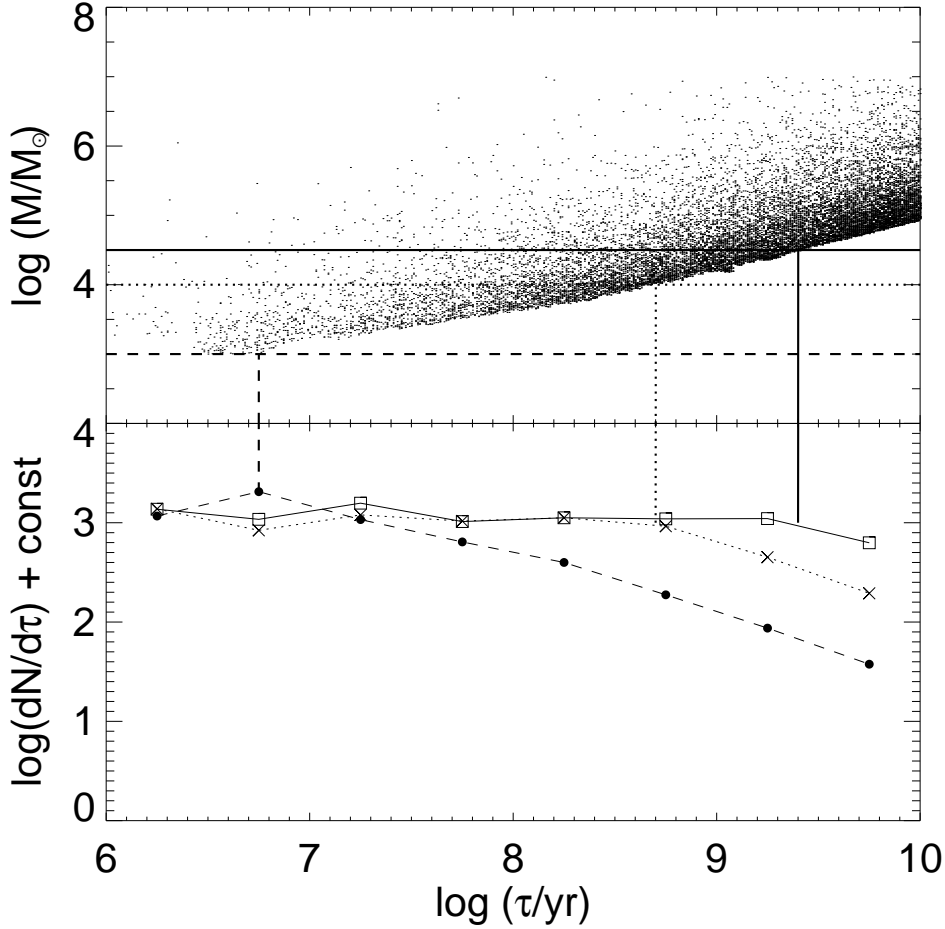}
\caption{A constant cluster formation model with no disruption
and an arbitrary V band detection limit that imposes a diagnoal
lower limit to the dataset (top panel).
The horizontal lines represent different cluster mass cuts that might
be used to study the sample.  The bottom panel shows that as the
mass limit used for plotting the age distribution is lowered
from $10^{4.5}$\msun\/ to $10^3$\msun\/, the age distribution goes
from flat to having an apparent bend.  The vertical lines show that
the location of the bend correlates with the age at which incompleteness
due to the magnitude limit occurs.
\label{dndtmasslimit}}
\end{figure}

\begin{figure}
\plotone{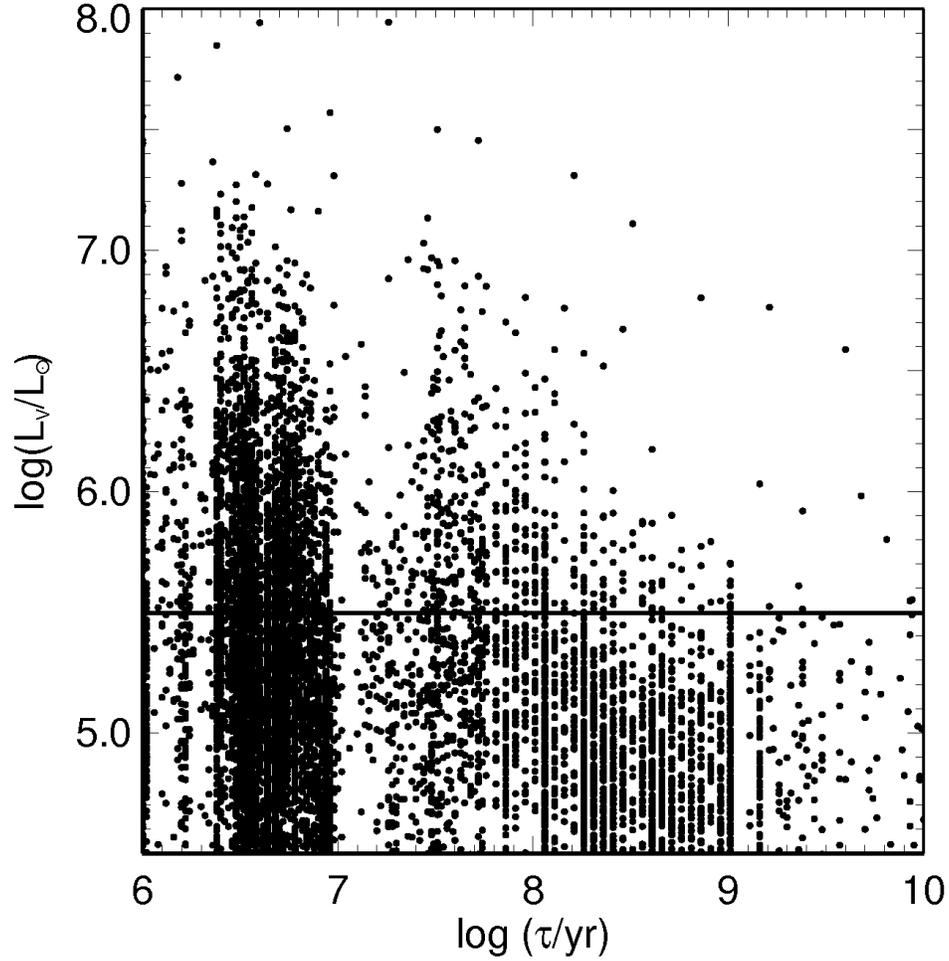}
\caption{Log luminosity versus log age diagram of star clusters in the
Antennae galaxies (as shown in Fall et al. 2005).  $L/L_{\odot}$ is the
extinction corrected
luminosity in the V-band.  The horizontal line at
$L=3\times10^5~L_{\odot}$ is the approximate upper limit for stellar
contamination.
\label{lt}}
\end{figure}

\begin{figure}
\epsscale{.75}
\plotone{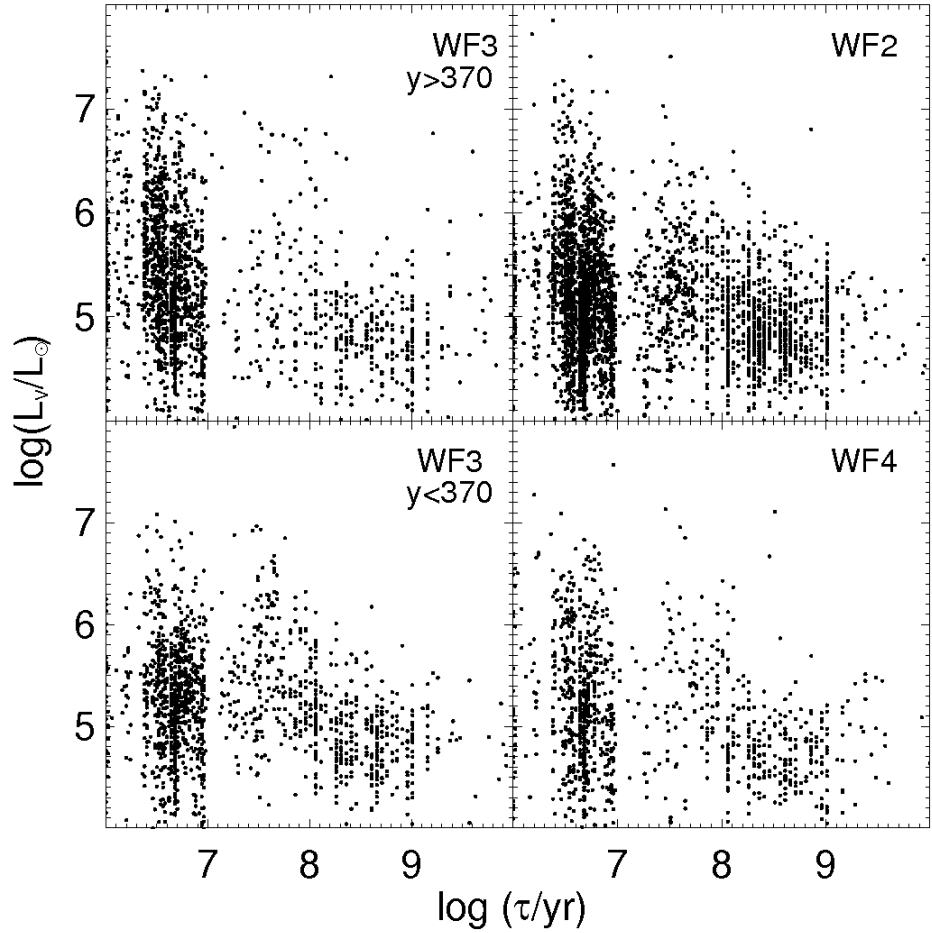}
\caption{Log luminosity versus log age of Antennae clusters
on different CCD chips. Note that the distribution on the WF3 chip
is broken into two groups; the dusty ``overlap region'' (Y$>$370),
and the
dust-free region (Y$<$370) with fewer very young clusters
\label{ltbyccd}}
\end{figure}

\begin{figure}
\epsscale{.75}
\plotone{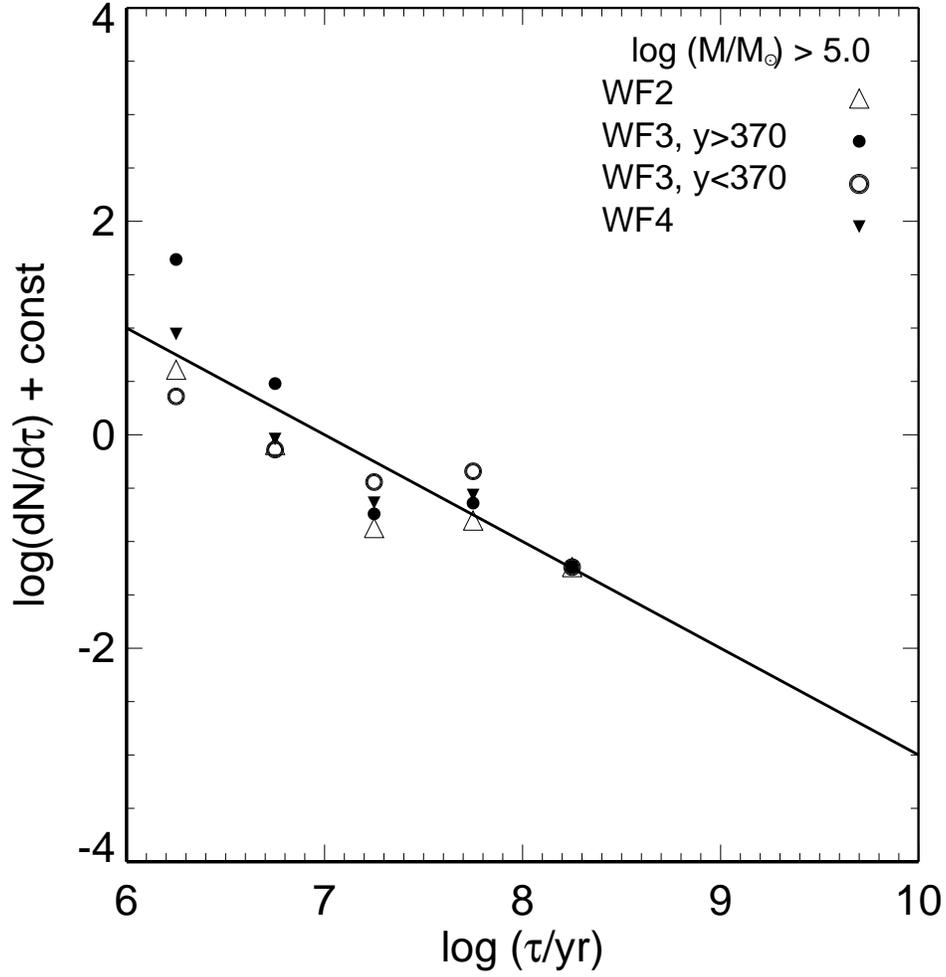}
\caption{Age distribution for the Antennae cluster system for
different CCD chips (normalized at log age = 8.25), as
defined in Figure 8.  
The solid line is $\gamma=-1$.
\label{dndtbyccd}}
\end{figure}

\begin{figure}
\epsscale{.75}
\plotone{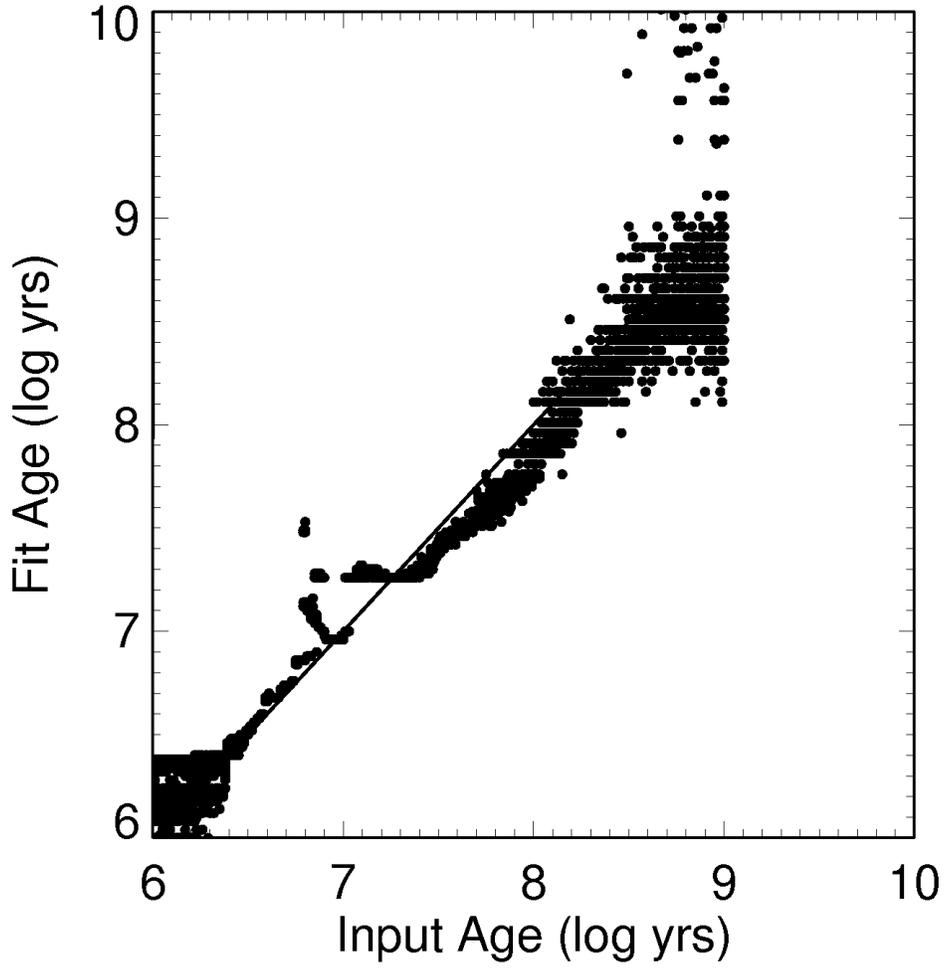}
\caption{Results of age-dating a synthetic cluster population (input
versus output age).  The solid line shows one-to-one
correspondence.  The technique is described in the text.
\label{ageinvsout}}
\end{figure}

\begin{figure}
\epsscale{0.4}
\plotone{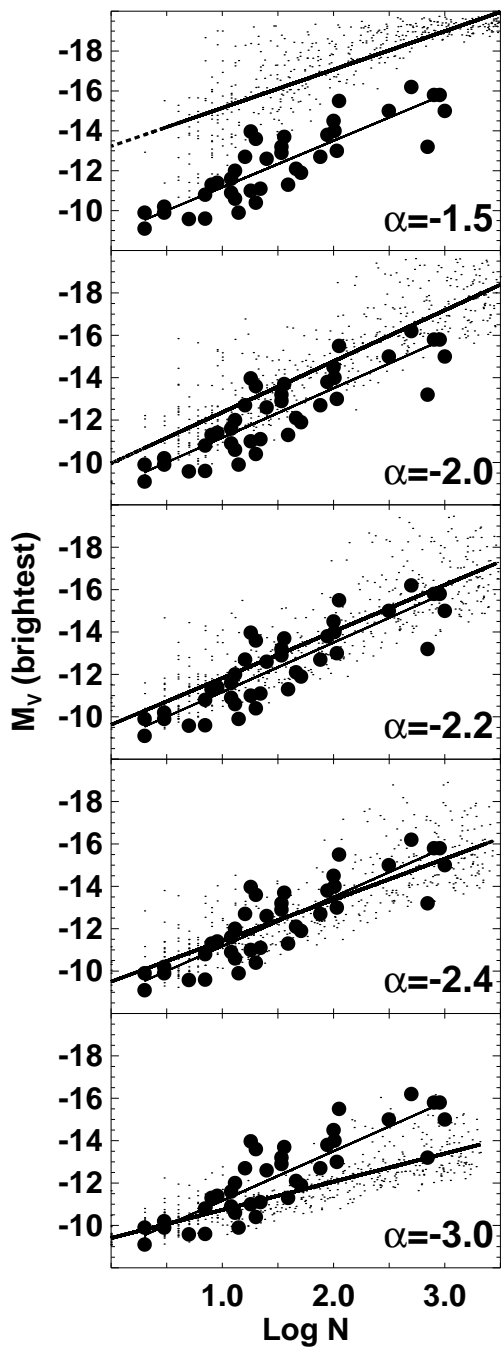}
\caption{Maximum observed cluster magnitude versus log of the total
number of clusters brighter than $M_V=-9$ from our 500 Monte
Carlo simulations (small dots). The simulations are compared with 
observations of cluster in 40 star-forming galaxies (filled circles), as described in the text.
The results are shown for five different values of $\alpha$, the
power law index for the cluster luminosity function.
\label{montestats}}
\end{figure}

\begin{figure}
\epsscale{0.5}
\plotone{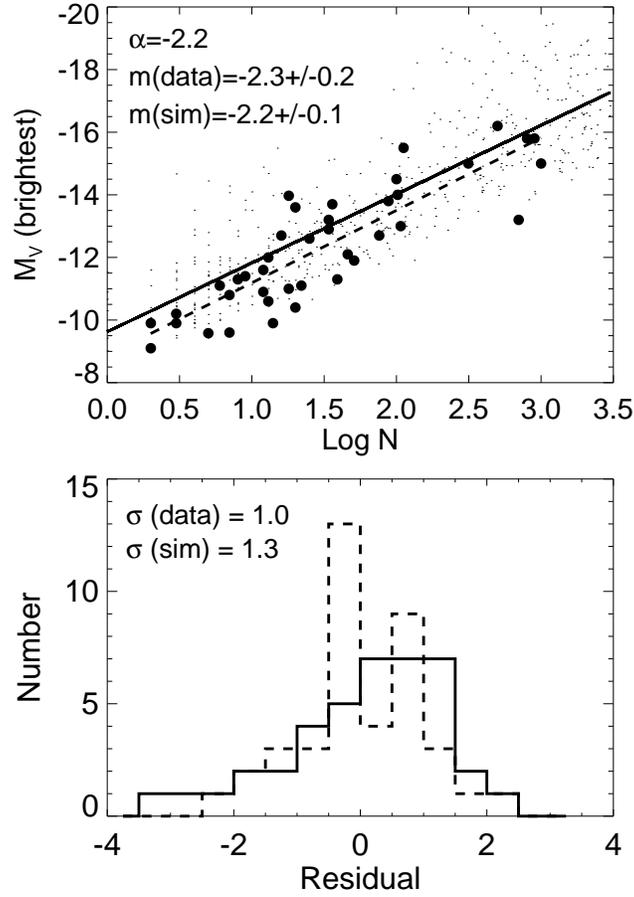}
\caption{The top panel reproduces the comparison between our
$\alpha=-2.2$ simulation (solid line) and data (dashed line) for the
M$_V$ (brightest) versus log N diagram, 
as shown in Figure~11.  The best fit slopes are given for both.
The bottom panel shows the dispersion ($\sigma$) of the residuals in
the data and simulations.  These are tabulated for all studied values
of $\alpha$ in Table~1.
\label{monteresid}}
\end{figure}

\begin{figure}
\epsscale{0.75}
\plotone{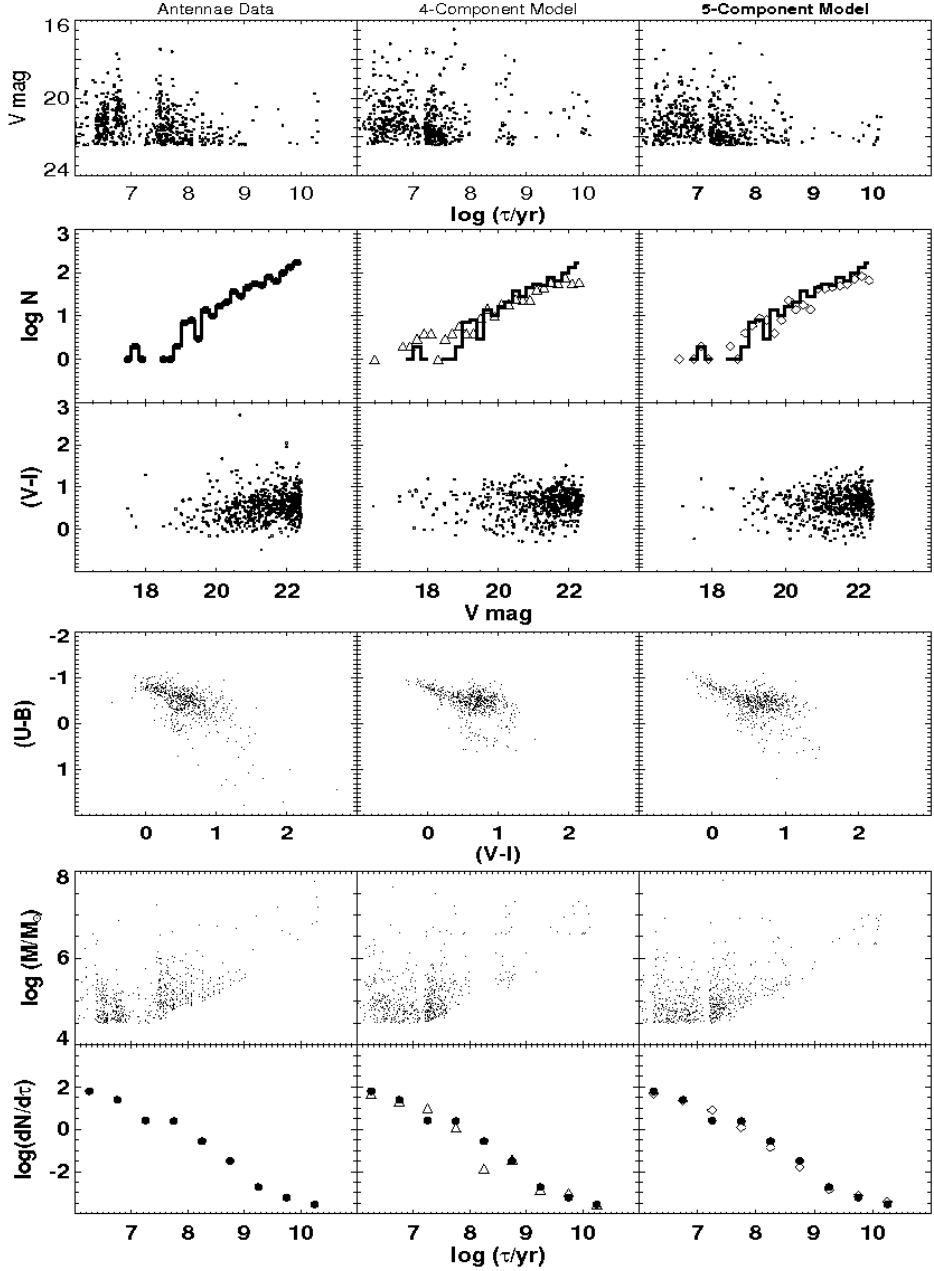}
\caption{The first column shows several observed and derived
diagnostics for the Antennae cluster system.  The second column
shows the same diagnostics for our original 4-component model for the Antennae, and
the third column shows the diagnostics for our final 5-component model (as described
in \S3.2).
\label{fig13}}
\end{figure}

\begin{figure}
\epsscale{0.8}
\plotone{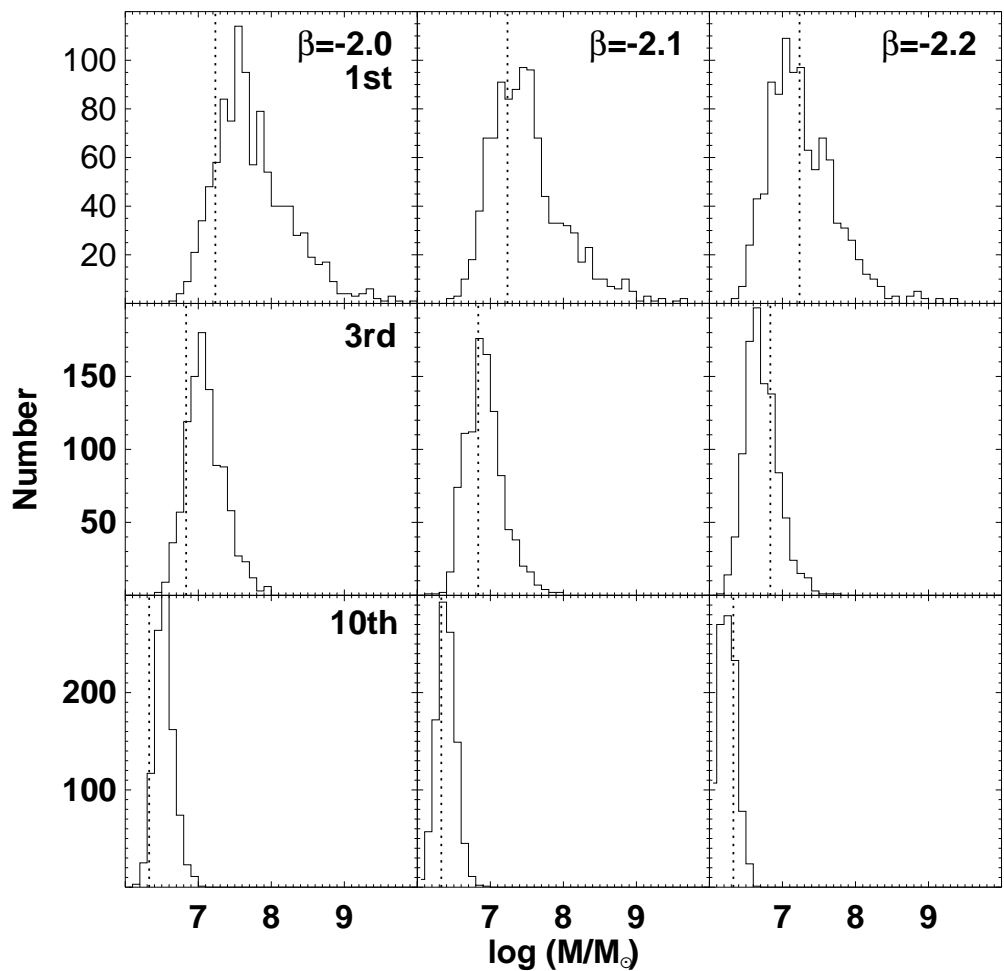}
\caption{Mass distributions for the first, third, and tenth most massive
cluster produced in 1000 Monte Carlo simulations.  The total
number of clusters used in each simulation is  matched to the
observed mass distribution in the range $4.5 \geq $ log $ M/M_{\odot}
\geq 5.5$ for the young Antennae cluster population ($10^6 \leq$ log
$\tau \leq 10^8$).  The three columns show simulations for $\beta$
values of $-2.0$, $-2.1$, and $-2.2$, and the dashed lines show the
measured mass for the Antennae cluster system. 
The match between simulations and measurements
show that a value of $\beta = -2.1$ provides good fits
without the need for a physical cutoff to the allowed
cluster masses in the Antennae.
\label{antmodel}}
\end{figure}

\begin{figure}
\epsscale{0.75}
\plotone{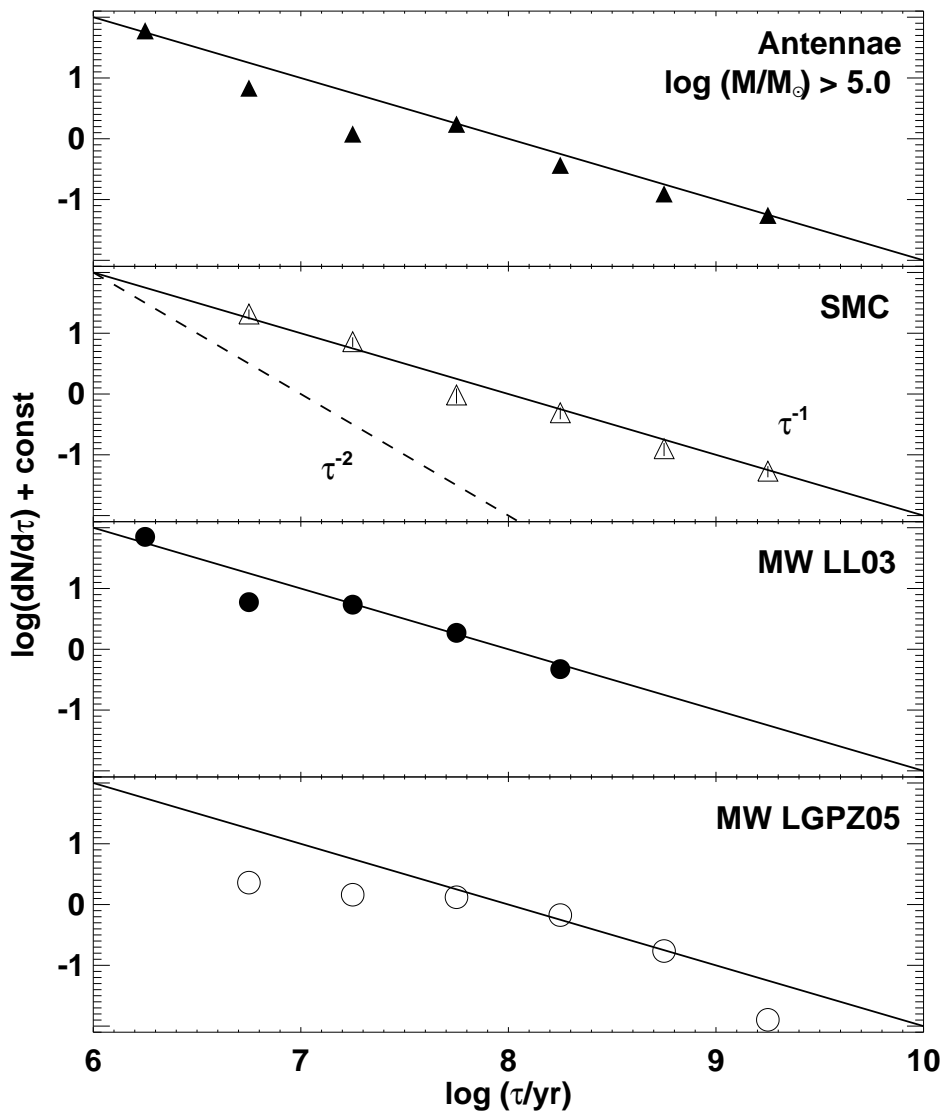}
\caption{The age distribution for clusters in the Antennae with masses
$\geq10^5~M_{\odot}$ (filled triangles); the SMC cluster population
from Rafelski \& Zaritsky (2005) but reinterpreted by
Chandar, Fall, \& Whitmore (2006); the embedded clusters in the Milky
Way (filled circles; Lada \& Lada 2003); and the open cluster sample
of Kharchenko et al. (2005) as studied by LGPZ05 (open circles).  
The solid lines
in each panel show $dN/d\tau \propto \tau^{\gamma}$, with $\gamma=-1$,
and the dashed line in the second panel  shows $\gamma=-2$ from Rafelski \& Zaritsky (2005) (see 
Chandar, Fall, \& Whitmore, 2006, for
a discussion).  
\label{fig:dndtlada}}
\end{figure}

\begin{deluxetable}{cccc}
\tablewidth{0pc}
\tablecaption{Results from $M_V$(brightest) versus log(N) Model
Simulations\label{table:stats}}
\tablehead{
\colhead{Powerlaw Index} &
\colhead{} & \colhead{} &
\colhead{Stdev of Residuals}\\
\colhead{$\alpha$} &
\colhead{Slope} &
\colhead{M$_V$(log[N]=1.5)} &
\colhead{$\sigma$}}
\startdata
$-1.5$ & $-$1.9 & $-$16.0 & 1.2\\
$-2.0$ & $-$2.4 & $-$13.6 & 1.3\\
$-2.2$ & $-$2.2 & $-$12.9 & 1.3\\
$-2.4$ & $-$1.9 & $-$12.4 & 1.2\\
$-3.0$ & $-$1.3 & $-$11.4 & 0.9\\
data\tablenotemark{a} & $-$2.3 & $-$12.4 & 1.0\\
\enddata
\tablenotetext{a}{The data for 40 galaxies have been collected
from the literature, as described in the text.}
\end{deluxetable}

\begin{deluxetable}{lccccccc}
\tablewidth{0pc}
\tablecaption{Final Results of Toy Model for the Antennae\label
{table:galprops}}
\tabletypesize{\footnotesize}
\tablehead{
\colhead{Galaxy} &
\colhead{Antennae} &
\multicolumn{6}{c}{5-Component Model}\\
\colhead{property} & \colhead{galaxy} &
\colhead{Total} &
\colhead{\#1 (const)} &
\colhead{\#2 (GC)} &
\colhead{\#3 (200 Myr)} &
\colhead{\#4 (1$-$100)} & \colhead{\#5 (1$-$10)}}
\startdata
stellar mass & 5e10 & 3.8e10 & 1.2e10 & 1.2e10 & 6.4e9 & 4.6e9 &
2.5e9\\
total $M_V$ & $-$21.7 & $-$22.2 & $-$18.7 & $-$18.1 & $-$20.1 & $-$21.1 & $-$21.6\\
\enddata
\tablerefs{\#1 = continuous creation over lifetime of the galaxy (1~Myr--13~Gyr)\\
\#2 = ancient Gaussian burst at $12.6\pm1$~Gyr (i.e., old globular cluster population)\\
\#3 = $200\pm100$~Myr Gaussian burst  (i.e. initial encounter)\\
\#4 = $1-100$~Myr continuous creation\\
\#5 = $1-10$~Myr continuous creation (recent burst observed on CCD WF3)}
\end{deluxetable}

\end{document}